**Population-average mediation analysis for zero-inflated count outcomes**

Andrew Sims PhD[1], D. Leann Long PhD[1], Hemant K. Tiwari PhD[1], Jinhong Cui MS[1], Dustin M. Long PhD[1], Todd M. Brown MD[2], MSPH, Melissa J. Smith PhD[1], Emily B. Levitan ScD[3]

1. University of Alabama at Birmingham, Department of Biostatistics

    Ryals Public Health Building (RPHB)

    1720 2nd Ave S

    Birmingham, AL 35294-0022

2. University of Alabama at Birmingham, Department of Medicine

    FOT 12

    1720 2nd Ave South

    Birmingham, AL 35294

3. University of Alabama at Birmingham, Department of Epidemiology

    Ryals Public Health Building (RPHB)

    1720 2nd Ave S

    Birmingham, AL 35294-0022

Summary

Mediation analysis is an increasingly popular statistical method for explaining causal pathways to inform intervention. While methods have increased, there is still a dearth of robust mediation methods for count outcomes with excess zeroes. Current mediation methods addressing this issue are computationally intensive, biased, or challenging to interpret. To overcome these limitations, we propose a new mediation methodology for zero-inflated count outcomes using the marginalized zero-inflated Poisson (MZIP) model and the counterfactual approach to mediation. This novel work gives population-average mediation effects whose variance can be estimated rapidly via delta method. This methodology is extended to cases with exposure-mediator interactions. We apply this novel methodology to explore if diabetes diagnosis can explain BMI differences in healthcare utilization and test model performance via simulations comparing the proposed MZIP method to existing zero-inflated and Poisson methods. We find that our proposed method minimizes bias and computation time compared to alternative approaches while allowing for straight-forward interpretations.

1. Introduction

Mediation analysis is increasingly used in medical and public health research to assess how much an intermediary variable, called a mediator, explains the relationship between an exposure and an outcome; however, these methods for count data with many zeroes are limited. Count outcomes with many zeroes occur often in public health research including healthcare utilization metrics (Liu et al., 2012), coronary artery stenosis (Orooji et al., 2022), surgery complication count (Wang et al., 2014), and dental caries (Preisser et al., 2012). When a count variable has many zeroes, standard count models like Poisson regression do not provide adequate inference on population effects (Lambert, 1992). In response the Zero-Inflated Poisson (ZIP) model was created to reduce biases for zero-inflated data (Mullahy, 1986). ZIP gives inference with respect to the probability of being an excess zero, that is zeroes beyond that which can be estimated by a Poisson distribution, and the mean of the non-excess zeroes (Mullahy, 1986). ZIP does not yield direct population mean estimates, which is necessary for population-based conclusions (Preisser et al., 2012). Although mean estimates can be derived from ZIP model, this formulation is conditional on covariates making interpretation challenging (Preisser et al., 2012).

Mediation analysis has become a popular tool for explaining causal relationships and providing inference on mechanistic pathways to intervention (VanderWeele, 2016b). Because mediation effect estimates rely on parameter estimates from the outcome model (VanderWeele, 2016b), using Poisson regression for a zero-inflated count outcome may give inadequate estimates of mediation effects(Lambert, 1992). While mediation methods for zero-inflated outcomes have been proposed utilizing the ZIP model (Wang and Albert, 2012; Cheng et al., 2018), the conditionality on covariate values makes interpretation of mediation effects difficult (Preisser et al., 2012). These methods also use the simulation-based approach to mediation, which can be computationally intensive for large datasets (Imai, Keele and Tingley, 2010; VanderWeele, 2016a).

In this paper we propose a new method for mediation with zero-inflated count outcomes by applying the Marginalized Zero-Inflated Poisson (MZIP) model (Long et al., 2014). MZIP marginalizes over the ZIP likelihood to directly obtain marginal mean estimates independent of covariate values (Long et al., 2014). Our mediation framework also uses the counterfactual approach to mediation, allowing for standard error estimation via bootstrapping and delta method for large datasets (Robins and Greenland, 1992; Pearl, 2001; VanderWeele, 2016a). By combining MZIP and the counterfactual approach to mediation, direct formulaic expressions of mediation effects can be derived that are independent of covariates in the outcome model, easy to interpret, and more computationally efficient than existing methods.

Section 2 reviews ZIP and MZIP. Section 3 reviews the counterfactual approach to mediation. Section 4 introduces mediation methodology for zero-inflated count outcomes using MZIP. Section 5 presents a simulation study, examining the properties of the proposed method. Section 6 presents an analysis using the proposed method with REasons for Geographic and Racial Differences in Stroke Study-Medicare linked data to evaluate how diabetes may explain BMI differences in inpatient care utilization in a Medicare population. A discussion follows in Section 7.

## 2. Zero-Inflated Models

### 2.1. Zero-Inflated Poisson Model

The ZIP distribution allows the count variable of interest, $Y_i, i = 1, \ldots n$, take the value of zero from a Bernoulli distribution with probability $\psi_i$ (the probability of being an excess zero) or arise from a Poisson distribution with probability $1 - \psi_i$ with mean $\mu_i$ (the mean of the non-excess zeroes) for each observation $i$ where $n$ is the sample size. To model the ZIP distribution, the ZIP model is expressed as:

$$logit(\psi_i) = \mathbf{Z}'_i \boldsymbol{\gamma}$$

$$\log(\mu_i) = \mathbf{Z}'_i \boldsymbol{\beta}$$

where $\boldsymbol{\gamma}$ and $\boldsymbol{\beta}$ are $(\rho \times 1)$ column vectors of parameters associated with the probability of being an excess zero and the mean of the non-excess zeroes respectively, $\mathbf{Z}_i$ is a $(\rho \times 1)$ vector of covariates for the $i$-th individual for the excess zero and non-excess zero mean components of ZIP, respectively, where $\rho$ is the number of parameters in the model. Often researchers are interested in the overall population effects instead of ZIP's latent class parameter estimates (Preisser et al., 2012). The overall population mean can be found in ZIP by the following formula $\nu_i = (1 - \psi_i)\mu_i = \frac{e^{\mathbf{Z}'_i \boldsymbol{\beta}}}{1 + e^{\mathbf{Z}'_i \boldsymbol{\gamma}}}$, but this derivation is not obvious to most researchers resulting in ZIP parameters being misinterpreted more often than not (Preisser et al., 2012). If one desired the overall population effect for a single covariate in the ZIP model, $z_{1i}$, this effect would be conditional on fixed values of other predictors making interpretation of effects challenging for population mean inference.

## 2.2. Marginalized Zero-Inflated Poisson Model

To directly obtain overall population effects when modeling data with excess zeroes, the Marginalized Zero-Inflated Poisson model was proposed by marginalizing over the likelihood of the ZIP model (Long et al., 2014). MZIP directly models $v_i$, the overall population mean. Specifically, MZIP model specifies:

$$logit(\psi_i) = \mathbf{Z}_i'\boldsymbol{\gamma}$$

$$log(v_i) = \mathbf{Z}_i'\boldsymbol{\alpha}$$

where, $\boldsymbol{\alpha}$ is a $(\rho \times 1)$ column vector of parameters associated with the overall population mean model, $\mathbf{Z}_i$ is a $(\rho \times 1)$ vector of covariates for the $i$-th individual for the excess zero and overall population mean components of MZIP, respectively and $\boldsymbol{\gamma}$ are parameters associated with the probability of being an excess zero. While $\gamma$ parameters have the same interpretations as in the ZIP model, exponentiating $\alpha$ parameters yield incidence rate ratio interpretations akin to Poisson and Negative Binomial models with respect to the overall population mean. Both components of MZIP can have different sets of covariates, but for simplicity we will assume the same covariates for both.

The marginalization of ZIP allows for overall mean parameters to be extrapolated directly from the model without dependencies on fixing values of covariates included in the model expression. These advantages make MZIP easier to interpret than ZIP when interest lies in the overall mean. In mediation analysis, researchers are often interested in evaluating the overall population effect of the relationship between the exposure and outcome operating through the mediator. Obtaining these overall mediation effects with ZIP would not be straightforward. The MZIP model has a distinct advantage in directly modeling overall mean.

## 3. Counterfactual Approach to Mediation

We build on the counterfactual approach to mediation because it allows for rapid derivation of standard errors via the delta method, can be easily extended to most generalized linear models, and can account for interaction effects (VanderWeele, 2016a). Assume $Y_i = Y$ is the observed value of the outcome, $M_i = M$ is the observed mediator, $X_i = X$ the observed exposure, and $\boldsymbol{C_i} = \boldsymbol{C}$ a vector of potential confounders (Figure 1). The counterfactual approach to mediation assumes there is no unmeasured confounding of the X-M, M-Y, and X-Y relationships and that no M-Y confounder is affected by X. Standard counterfactual notation is as follows: Assume the exposure $X$ takes two levels, $x$ and $x^*$, and without loss of generality let X be binary where $x = 1$ and $x^* = 0$ are considered treatment and control groups respectively.

- $Y(x, m) = Y_{xm}$ is the counterfactual outcome that would be observed if an individual's treatment were set to $X = x$ and their mediator were set to $M = m$

- $M_{x*}$: is the counterfactual mediator that would be observed if an individual's treatment were set to $X = x^*$

- $Y(x, M_{x*}) = Y_{xM_{x^*}}$ is the counterfactual outcome for an individual if they received treatment, but the mediator was set to the value it would have taken naturally under control

We also adopt the consistency assumption, that is when $X = x$ the observed outcome $Y$ and mediator $M$ are equal to their counterfactual values $Y(x)$ and $M(x)$(VanderWeele and

Vansteelandt, 2009). In addition, when $X = x$ and $M = m$ the observed outcome $Y$ is equal to the counterfactual outcome $Y(x, m)$(VanderWeele and Vansteelandt, 2009).

To estimate mediation effects two regression models are fitted. One regression model uses the exposure, mediator, and any covariates on the outcome and the second regression model uses the exposure and covariates on the mediator.

$$g(E[Y|X = x, M = m, C = c]) = g(E[Y|x, m, c]) = \tau_0 + \tau_1 x + \tau_2 m + \tau_3 xm + \tau_4' c$$

$$h(E[M|X = x, C = c]) = h(E[M|x, c]) = \theta_0 + \theta_1 x + \theta_2' c$$

where $g$ and $h$ are the link functions associated with the outcome and mediator model respectively and $\tau$ and $\theta$ are the parameter estimates associated with the outcome and mediator model respectively.

The following significant quantities can be estimated using expected values and parameter estimates from the above equations. Prominent effects such as the total effect (TE), natural direct effect (NDE), natural indirect effect (NIE), and controlled direct effect (CDE) can be computed based on closed form expressions, but the complexity of these expressions depend on the link function of the outcome model (Robins and Greenland, 1992; Pearl, 2001; VanderWeele, 2016a). Below formulas for each effect are shown.

$$NDE = E(Y_{xM_{x^*}} - Y_{x^*M_{x^*}}|c) = \sum_m \{E[Y|x, m, c] - E[Y|x^*, m, c]\}P(m|x^*, c)\}$$

$$NIE = E(Y_{xM_x} - Y_{xM_x^*}|c) = \sum_m E[Y|x, m, c](P(m|x, c) - P(m|x^*, c))$$

$$CDE(m) = E(Y_{xm} - Y_{x^*m}|c) = E(Y|x, m, c) - E(Y|x^*, m, c)$$

$$TE = E(Y_{xM_x} - Y_{x^*M_{x^*}}|\boldsymbol{c}) = \sum_m E[Y|x,m,\boldsymbol{c}]P(m|x,\boldsymbol{c}) - \sum_m E[Y|x^*,m,\boldsymbol{c}]P(m|x^*,\boldsymbol{c})$$

Interpretations of these effects are tied to link functions used in the outcome model (i.e.., $g(Y)$). If an identity link function is used effects will be on a risk difference (RD) scale, alternatively if a log-link is used effects will be on a ratio scale such as incidence rate ratio (IRR) also called incidence density ratios for Poisson outcomes. It is also common to compute effects on a RD scale regardless of the link function of the outcome model. On a difference scale the total effect is equivalent to the sum of NDE and NIE, but on a ratio scale, the NDE and NIE are multiplied to obtain the TE [$IRR^{TE} = IRR^{NIE}(IRR^{NDE})$] (Valeri and VanderWeele, 2013; VanderWeele, 2016a). Note that when there is no interaction effect between the exposure and mediator $CDE = NDE$.

Standard errors for mediation effects can be computed through computationally intensive bootstrapping techniques (Lockwood and MacKinnon, 1998) or closed form expressions via delta method (VanderWeele, 2016a).

4. Mediation with Zero-Inflated Outcome using MZIP

Methods have been previously proposed for mediation with zero-inflated outcomes using ZIP and the simulation-based approach to mediation (Wang and Albert, 2012; Cheng et al., 2018). The simulation-based approach is similar to the counterfactual approach except after fitting the outcome and mediator models, parameters are then simulated from their sampling distribution based on a normal distribution with mean equal to the fit model parameter estimate and variance equal to the fit model estimated asymptotic covariance matrix (King, Tomz and Wittenberg, 2000; Imai et al., 2010). The simulation-based approach has been documented in R software, but its use can be computationally intensive for large datasets and it only provides direct and indirect effects on a risk difference scale (Tingley et al., 2014). Using ZIP to obtain overall mediation effects results in formulas that are conditional on covariates in complicated ways, making interpretation difficult as they vary by fixed values of covariates. This problem is exacerbated when computing effects on a difference scale as formulas lack convenient simplification.

For illustration, let

$$logit(\psi_i|x, m, \mathbf{c}) = \gamma_0 + \gamma_1 x + \gamma_2 m + \boldsymbol{\gamma_4}'\mathbf{c}$$

$$\log(\mu_i|x, m, \mathbf{c}) = \beta_0 + \beta_1 x + \beta_2 m + \boldsymbol{\beta_4}'\mathbf{c}$$

$$E(M|x, \mathbf{c}) = \theta_0 + \theta_1 x + \boldsymbol{\theta_4}'\mathbf{c}$$

where modeling of $\psi$ and $\mu$ corresponds to the equations above. We let the mediator, $M$, be a continuous variable modeled by a linear regression. The NDE for the overall population mean on the risk difference scale is

$$NDE_{ZIP} = \left[\frac{\exp(\beta_0 + \beta_1 x + \beta_2 m + \boldsymbol{\beta_4'c})}{1 + \exp(\gamma_0 + \gamma_1 x + \gamma_2 m + \boldsymbol{\gamma_4}'\mathbf{c})} - \frac{\exp(\beta_0 + \beta_1 x * + \beta_2 m + \boldsymbol{\beta_4'c})}{1 + \exp(\gamma_0 + \gamma_1 x^* + \gamma_2 m + \boldsymbol{\gamma_4}'\mathbf{c})}\right] P(M|x*)$$

and on the ratio scale is

$$IRR_{ZIP}^{NDE} = e^{\beta_1(x-x*)} \frac{1 + \exp(\gamma_0 + \gamma_1 x^* + \gamma_2 m + \boldsymbol{\gamma_4}'\mathbf{c})}{1 + \exp(\gamma_0 + \gamma_1 x + \gamma_2 m + \boldsymbol{\gamma_4}'\mathbf{c})} P(M|x*)$$

These expressions of NDE would be challenging to derive because $P(M|x^*)$ cannot be easily placed within the expression. Note that the NDE on both scales depends on covariates in both the excess zero and the Poisson component of ZIP and the mediator model. This is problematic because there is no unique NDE as NDE will vary by covariate values. Often mean values of covariates are used to approximate marginal means, but when covariates follow a skewed distribution, this may not result in estimates generalizable to the population. This problem holds for NIE.

We propose utilization of MZIP instead of ZIP to directly obtain overall population mean parameters to simplify derivation and interpretation of mediation effects. Specifying the MZIP outcome model as

$$logit(\psi_i|x,m,\boldsymbol{c}) = \gamma_0 + \gamma_1 x + \gamma_2 m + \boldsymbol{\gamma_4}'\boldsymbol{c}$$

$$log(v_i|x,m,\boldsymbol{c}) = \alpha_0 + \alpha_1 x + \alpha_2 m + \boldsymbol{\alpha_4}'\boldsymbol{c}$$

Yields

$$NDE_{MZIP} = e^{\alpha_0 + \alpha_4'c + (\theta_0 + \theta_1 x* + \theta_4'c)\alpha_2 + \frac{\sigma^2 \alpha_2^2}{2}}(e^{\alpha_1 x} - e^{\alpha_1 x*})$$

$$NIE_{MZIP} = e^{\alpha_0 + \alpha_1 x + (\theta_0 + \theta_4'c)\alpha_2 + \frac{\sigma^2 \alpha_2^2}{2} + \alpha_4'c}(e^{\theta_1 x \alpha_2} - e^{\theta_1 x* \alpha_2})$$

On a risk difference scale, NDE and NIE for MZIP will still be conditional on covariates for both outcome and mediator models but doesn't depend on parameters from the zero-inflated component of MZIP. However, on a ratio scale we obtain:

$$IRR_{MZIP}^{NDE} = \frac{e^{\alpha_0 + \alpha_1 x + \alpha_2(\theta_0 + \theta_1 x* + \theta_4'c) + \alpha_4'c}}{e^{\alpha_0 + \alpha_1 x* + \alpha_2(\theta_0 + \theta_1 x* + \theta_4'c) + \alpha_4'c}} = e^{\alpha_1(x - x*)}$$

$$IRR_{MZIP}^{NIE} = \frac{e^{\alpha_0 + \alpha_1 x + \alpha_2(\theta_0 + \theta_1 x + \theta_4'c) + \alpha_4'c}}{e^{\alpha_0 + \alpha_1 x + \alpha_2(\theta_0 + \theta_1 x* + \theta_4'c) + \alpha_4'c}} = e^{\alpha_2 \theta_1(x - x*)}$$

These quantities are independent of covariate values. Using IRR formulations of mediation effects with MZIP, we obtain unique effect estimates that are independent of covariate values making interpretation consistent across all potential covariate combinations.

We also extend derivations of mediation effects to cases with exposure-mediator interactions. For a continuous mediator, the models can be expressed by

$$\log(v_i|x,m,\boldsymbol{c}) = \widetilde{\alpha_0} + \widetilde{\alpha_1}x + \widetilde{\alpha_2}m + \widetilde{\alpha_3}xm + \widetilde{\boldsymbol{\alpha_4}}'\boldsymbol{c}$$

$$logit(\psi_i|x,m,\boldsymbol{c}) = \widetilde{\gamma_0} + \widetilde{\gamma_1}x + \widetilde{\gamma_2}m + \widetilde{\gamma_3}xm + \widetilde{\boldsymbol{\gamma_4}}'\boldsymbol{c}$$

$$E(M_i|x,\boldsymbol{c}) = \theta_0 + \theta_1 x + \boldsymbol{\theta_4}'\boldsymbol{c}$$

NDE, NIE, and CDE incidence rate ratios can then be derived by the following formulas

$$IRR_{MZIP}^{NDE} = e^{[\widetilde{\alpha_1}+\widetilde{\alpha_3}(\theta_0+\theta_1 x*+\theta_4 c+\widetilde{\alpha_2}\sigma^2)](x-x*)+\frac{1}{2}\widetilde{\alpha_3}^2\sigma^2(x^2-x*^2)}$$

$$IRR_{MZIP}^{NIE} = e^{(\widetilde{\alpha_2}\theta_1+\widetilde{\alpha_3}\theta_1 x)(x-x*)}$$

$$IRR_{MZIP}^{CDE} = e^{(\widetilde{\alpha_1}+\widetilde{\alpha_3}m)(x-x*)}$$

where $\sigma^2$ is the variance of the linear mediator model, which can be estimated by taking the residual sum of squares and dividing by the difference in the sample size and the number of parameters in the model (VanderWeele, 2016a). Adding the interaction term makes it impossible to avoid NDE being conditional on covariates, but use of MZIP minimizes this dependency. NIE is the most desired quantity in mediation analysis and even with an interaction effect, $RR^{NIE}$ using MZIP can be obtained unconditional on covariates making interpretation of mediation effects more straightforward. Proof of these derivations can be found in Web Appendices A and B and standard errors can be computed using bootstrapping or delta method derivations (Web Appendix C). We have also computed effects for cases with binary mediators (Web Appendices D and E) along with their respective delta method error derivations (Web Appendix F). Poisson models are not robust to overdispersion, but use of robust standard errors for MZIP has been shown to provide adequate coverage for over-dispersed zero-inflated count outcomes (Preisser et al., 2016). Because delta method uses the covariance matrix from MZIP, we can use the robust covariance matrix in MZIP to obtain robust standard errors for mediation effects when overdispersion is a concern (Tibshirani, 1996).

## 5. Simulation

Properties and performance of the proposed mediation method using MZIP are examined in this section. Ratio scale NDE and NIE estimates for the proposed method using a MZIP outcome model are compared to cases using a Poisson outcome model using the counterfactual approach to mediation. These comparisons are made based on median percent bias, coverage probability, power, and median standard error (MSE) for both delta method and bootstrapped standard errors. 1000 resamples are used for bootstrapping.

The exposure, $X$, is simulated from a random Bernoulli distribution with probability 0.5. A single covariate with a skewed distribution, $C$, is simulated following a $\chi_2^2$ distribution. A continuous mediator $M$ is simulated based on a random normal distribution such that

$$M = \theta_0 + \theta_1 x + \theta_4 c + \epsilon$$

where $\epsilon \sim N(0, \sigma^2)$ and $\sigma^2 = 3$. Let $\mathbf{Z}$ be a $(n \times 4)$ matrix of simulated exposure, mediator, and covariate values with an intercept vector are merged, where $\mathbf{Z} = (\mathbf{1}, x, m, c)$ and $n$ is the simulated sample size. The zero-inflated outcome $Y$ is then simulated with MZIP model specifications by

$$\psi \sim Bernoulli\left(\frac{\exp(Z\gamma)}{1 + \exp(Z\gamma)}\right)$$

$$\mu \sim Poisson(\exp(Z\alpha + \log(1 + Z\gamma)))$$

where outcome $Y = \nu = (1-\psi)\mu$.

Sample sizes of 200, 600, and 1000 are considered for each scenario discussed with 5000 iterations each. While more scenarios were considered (Web Appendix G), this discussion will focus on two scenarios on both a ratio (Table 1). Scenario 1 focuses on zero-inflated models with a high impact of the exposure on the probability of being an excess zero ($\gamma_1$) while Scenario 2 increases the mediator's effect on the probability of being an excess zero ($\gamma_2$) from Scenario 1 to see how this increase affects NIE and NDE estimation. We hypothesize that the Poisson model will perform worse under these conditions because it does not account for differential effects of covariates on the probability of being an excess zero. The simulated outcome distribution contains about 75% and 80% zeroes in the control group and 45% and 55% in the treatment group for Scenarios 1 and 2 respectively. For our simulation study, the mean value of 2 was used for the covariate to approximate marginal estimates for the $\chi_2^2$ covariate.

The proposed MZIP model is compared to a Poisson outcome model using the counterfactual approach to mediation for Scenario 1 (Table 2) and Scenario 2 (Table 3). Though the percent relative bias for both Poisson and MZIP outcome models decreases as sample size increased, the Poisson outcome model produced severely biased results even at a sample size of 1000. This was exacerbated when raising the zero-inflation from the mediator in Scenario 2 reinforcing results from literature beyond mediation analysis (Lambert, 1992; Long et al., 2014).

Corresponding to this increased bias, the Poisson outcome model also had abysmal coverage particularly when using delta method standard errors with coverage of 17.1% for NDE and 51.2% for the NIE for sample size 1000 in Scenario 2. Bootstrap standard errors for Poisson were slightly below the nominal 95% in Scenario 1 but were poor in Scenario 2. Based on the MSE results, both delta method and bootstrapped standard errors for Poisson significantly underestimated the intrinsic variance, explaining the poor coverage estimates. Not only does Poisson regression yield inaccurate estimates of NIE and NDE, but it also fails to adequately estimate variance for zero-inflated outcomes. There was a significant difference in power between the two variance estimation techniques for the Poisson outcome model. Power via the delta method tended to be much higher than via bootstrapped because of the extremity in the delta method in underestimating variance.

In comparison, coverage for MZIP outcome models was near the nominal 95% for n=600 and n=1000 and only slightly below 95% at n=200 for NIE. Power was similar for both the delta method and bootstrap methods for MZIP outcome models. MSE estimates for both variance estimation techniques also approximated the nominal value for all cases. This implies that MZIP outcome models perform well in estimating NDE, NIE, and their respective variances. Both the delta method and bootstrapping are adequate in estimating errors for zero-inflated count outcomes. As bootstrapping tends to be computationally intensive, the convergence of variance estimation with delta method at smaller sample sizes means reliance on computationally intensive methods may not even be necessary when using MZIP.

Poisson regression does not account for differential effects of the mediator in the probability of being an excess zero which explains why results for Poisson outcomes in Scenario 2 had increased bias and decreased coverage for both NIE and NDE estimates. Other scenarios were considered (Web Appendix G) accounting for changes in zero-inflation and overall mean that amplify the importance of zero-inflation in performance for the Poisson model. Scenarios for binary mediators were also considered. These scenarios performed similarly to the case with continuous mediators with Poisson having biased estimates with poor coverage (Web Appendix H). We also considered simulation scenario based on Scenario 2 with an over-dispersed zero-inflated count outcome (Web Appendix I). In this scenario delta method variance estimates slightly underestimated variance when using the proposed method. Bootstrapping or use of a robust covariance matrix for MZIP seems to still provide nominal coverage and ideal variance estimation despite over-dispersion.

In addition, we considered simulations on a difference scale comparing to the exiting simulation-based approach using ZIP (Cheng et al., 2018). However, ZIP and MZIP model assumptions are slightly different and because our simulations were built upon an MZIP model, the ZIP method will naturally be more biased (Liu et al., 2018). This information is shown in Web Appendix G; however, note that the results in no way imply that the proposed method or existing method are superior to each other in terms of bias. We did, however, compare the methods in terms of average computation speed. We found that at a sample size of 1000 the proposed method using the delta method took 0.3 seconds compared to 300 seconds for bootstrap. Whereas the existing method using quasi-Bayesian standard errors took about 44x longer than the proposed method using delta method. Comparisons of the proposed method using MZIP and Poisson were equivalent on a difference scale (Web Appendix J).

## 6. Illustrative Application

Obesity, defined as body mass index (BMI) >30 kg/m$^2$, is associated with an increased risk of various comorbidities such as cardiovascular disease, diabetes, and obstructive sleep apnea (Bray et al., 2018). Obesity itself is associated with higher health care costs and inpatient utilization (Wang et al., 2005; Cecchini and Sassi, 2015; Peterson and Mahmoudi, 2015; Musich et al., 2016). It also stands that these comorbidities that are common in those who are obese could also be associated with higher health care utilizations and costs (American Diabetes Association, 2018). Using the proposed novel mediation method, we will analyze how differences between BMI categories in number of inpatient care visits, a zero-inflated count variable, may be explained by the presence of diabetes.

The REasons for Geographical and Racial Differences in Stroke (REGARDS) study is an ongoing national cohort of 30,239 US adults aimed at identifying factors that explain geographical and racial disparities in stroke (Howard et al., 2005). Black and White individuals over age 45 were enrolled between 2003-2007 oversampling Black adults and people from the Stroke Belt and Stroke Buckle, a region in the Southeast US with higher stroke incidence (Howard, 1999). Demographic data were collected via telephone interviews, and physical measurements such as BMI were assessed during an in-home examination. A subset of REGARDS participants were able to be linked to Medicare claims data for healthcare utilization metrics (Xie et al., 2016). For this application, 9726 participants are included from this Medicare claims cohort with an average age of 72 years, 51% women, and 35% Black. Specifically, researchers are interested in whether BMI differences in inpatient care (during the 2 years following baseline) can be explained by having diabetes, defined as fasting blood glucose≥126 mg/dL, non-fasting blood glucose ≥200 mg/dL, or prescribed diabetes medication.

BMI is categorized into 5 categories, 18.5-<25 kg/m$^2$, 25-<30 kg/m$^2$, 30-<35 kg/m$^2$, 35-<40 kg/m$^2$, ≥40 kg/m$^2$. Control was made for potential confounders including age, sex, race, geographical region (Stroke Belt, Stroke Buckle), education level (at least some college vs. no college), physical activity (none, 1-3 times/week, 4 or more per week) and smoking status

(current, past, never). Inpatient care is defined as the number of inpatient care visits within 2 years from REGARDS study enrollment and is over 60% zeroes (Figure 2). Analysis was done using the proposed MZIP mediation method and a Poisson outcome model.. Results are displayed in Table 4. Bootstrap standard errors can be found in Web Appendix K.

First, we assess the total effect or the relationship between BMI and inpatient care using MZIP, we found that overall, as BMI increased there was an increased incidence of inpatient care visits. When compared to BMI 18.5-<25 kg/m$^2$ (normal BMI) those with BMI 30-<35 kg/m$^2$ ($IRR^{TE} = 1.22, CI: 1.10, 1.35$), 35-<40 kg/m$^2$ ($IRR^{TE} = 1.24, CI: 1.07, 1.44$) , ≥40 kg/m$^2$ ($IRR^{TE} = 1.46, CI: 1.23, 1.73$) had a statistically significant higher incidence of inpatient care visits (Table 4). However, after factoring in the mediator, diabetic status, we no longer observed a statistically significant exposure-mediator relationship for BMI 30-<35 kg/m$^2$ ($IRR^{NDE} = 1.10, CI: 0.99, 1.22$), and 35-<40 kg/m$^2$ ($IRR^{NDE} = 1.07, CI: 0.93, 1.24$), but BMI ≥40 kg/m$^2$ ($IRR^{NDE} = 1.23, CI: 1.03, 1.46$) had a statistically significant NDE (Table 4). Comparing BMI ≥40 kg/m$^2$ to the normal group, a significant portion of the total effect of the relationship between BMI and inpatient care visits operated through having diabetes ($IRR^{NIE} = 1.19, CI: 1.14, 1.23$) (Table 4). We also observed significant indirect effects for each BMI group implying there is a significant pathway between BMI categories and inpatient care operating through having diabetes. In summary we found that those with BMI ≥40 kg/m$^2$ had a 46% increased incidence of inpatient care visits than those with BMI 18.5-<25 kg/m$^2$; when factoring in diabetes status this increased incidence reduced to 23% with the remainder being attributable to diabetes status. We could conclude that about 50% of the difference in inpatient care between those with BMI≥ 40 and normal BMI can be explained by diabetes status. Overall, it seems that more of the BMI differences in inpatient care utilization are operating through having diabetes than is truly attributable to BMI.

Both MZIP and Poisson methods yielded identical interpretations. NIE and NDE point estimates were similar for the MZIP and Poisson outcome models, but confidence intervals were narrower when using Poisson outcome models (Web Appendix L), specifically for NDE, which, as shown in the simulation results, can have detrimental effects on scientific conclusions. Given the differences between effect estimates using model-based and robust errors for MZIP, it seems that there may be over-dispersion with model-based MZIP estimates underestimating variance. Use of robust standard error alleviates some of the burden of over-dispersion and had confidence intervals akin to bootstrap confidence intervals using the proposed method (Web Appendix K).

We also completed this analysis on a difference scale (Web Appendix J). The most noticeable difference between ZIP and MZIP methods was computation speed. The simulation-based approach using ZIP with quasi-Bayesian errors took 3.4 times longer than the proposed method with delta method errors (300 vs 60 seconds) while bootstrapping took over 38 hours.

## 7. Discussion

A novel mediation method for zero-inflated count outcomes using a MZIP outcome model in the counterfactual framework to mediation was proposed. This method lessens the conditionality of mediation effects on covariates, increases computation efficiency, and gives unbiased estimates of mediation effects and their respective standard errors. The proposed method is compatible with continuous and binary mediators and with exposure-mediator interactions.

The simulation study discussed in Section 5 demonstrated the danger of ignoring zero-inflation for a count outcome. Use of a Poisson outcome model gave highly biased estimates of direct and indirect effects and severely underestimated their respective variances. These results confirm previous study results that Poisson regression should not be used for zero-inflated count variables because they may produce inaccurate and unreliable scientific conclusions (Lambert, 1992; Long et al., 2014). The simulation study showed that the proposed method using MZIP gave unbiased estimates of NIE and NDE and their respective variances. R code to implement the proposed method is available through the Comprehensive R Archive Network (CRAN) (Sims et al., 2023).

Bootstrapping is the standard method of estimating variance for mediation effects, but this can be computationally intensive for large datasets like the REGARDS example (Lockwood and MacKinnon, 1998; VanderWeele, 2016a). Our proposed method allows for variance to be estimated via closed form expressions using the delta method. Our simulation study showed that regardless of sample size delta method estimates performed similarly to bootstrapped estimates. These closed-form expressions of variance can be computed in under a second, minimizing analytic time while still obtaining reliable estimation. Both variance estimation techniques gave reliable estimates and can be utilized in the R package.

While other methods exist for mediation with zero-inflated outcomes, these methods give estimates that are conditional on fixed values of covariates (Cheng et al., 2018). This is problematic when estimates (and potentially inference) vary depending on covariate values, making inference on population-average effects challenging and easily misinterpreted which may result in misleading conclusions. Using the mean value of covariates one can approximate marginal effect estimates; however, the proposed method simplifies this process by giving marginal mean effect estimates for the zero-inflated outcome when effects are assessed on a ratio scale. Our simulation and application showed that compared to the existing method, the method proposed in this manuscript is more computationally efficient. For large datasets the proposed method using MZIP with delta method standard errors can save researchers a significant amount of time in the statistical analysis portion of manuscript preparation while obtaining more straight-forward effects than existing methods.

While we have only considered continuous and binary mediators here, future work will seek to expand this methodology to other classes of mediator variables such as count and time-to-event. Future work will also seek to expand this methodology to other zero-inflated variable models such as marginalized zero-inflated binomial(Preisser et al., 2016), and cases with multiple mediators.

## 8. Conclusion

This paper proposed a mediation method for count outcomes with excess zeroes using MZIP to obtain marginal inference of the exposure-outcome and mediator-outcome relationships. Traditional count models, like Poisson, give biased parameter estimates that underestimate variance, while other past mediation methods for zero-inflated outcomes are computationally intensive, challenging to interpret due to estimates being conditional on covariate values, and can be unstable when the probability of excess zero attributable to the mediator is large. The proposed method using MZIP gives unbiased and marginal NDE and NIE estimates, reliable standard error coverage, and lessens the computational burden of statistical analysis. While the illustrative application focused on healthcare utilization, this analysis can be used to explore a wide variety of applications involving zero-inflated count outcomes.

Sources of Funding

This research project is supported and co-funded by the National Institute of Neurological Disorders and Stroke and the National Institute on Aging (cooperative agreement U01 NS041588). Additional funding was provided by Amgen, Inc. Representatives of Amgen reviewed the manuscript but were not involved in the collection, management, analysis, or interpretation of the data or in the decision to publish this work.

This project is supported by a National Heart, Lung, and Blood Institute (NHLBI) pre-doctoral training fellowship (T32 HL155007). Its contents are solely the responsibility of the authors and do not necessarily represent the official views of the NHLBI or NIH.

Acknowledgements

We thank the other investigators, the staff, and the participants of the Reasons for Geographic and Racial Differences in Stroke study for their valuable contributions. A full list of participating Reasons for Geographic and Racial Differences in Stroke investigators and institutions can be found at http://www.regardsstudy.org.

Supporting Information

Web Appendices, Tables, and Figures referenced in Sections 4-6 are available with this paper at the Biometrics website on Wiley Online Library. REGARDS data is available to qualified researchers upon approval of a research proposal and data use agreement. Details are available at regardsstudy.org. The `maczic` R package available on CRAN was used for the existing methodology for mediation with zero-inflated count outcomes (Cheng, Guo and Cheng, 2021) (https://cran.r-project.org/web/packages/maczic/index.html). The `mzipmed` R package was developed for implementation of the methods described in this manuscript (https://cran.r-project.org/web/packages/mzipmed/index.html).

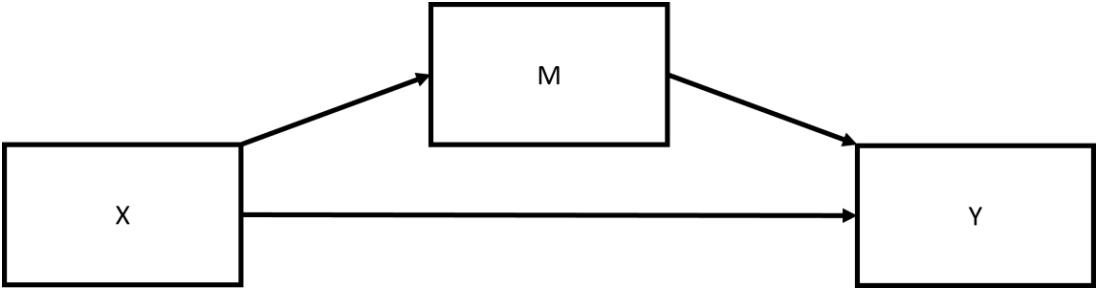

*Figure 1: Pathways of a Standard Mediation Analysis with exposure, X, mediator, M, and outcome, Y.*

**Table 1: Scenarios considered in simulation study of mediation with zero-inflated count outcome***

| Scenario | MZIP Parameters | Mediator Model | NDE | NIE |
|---|---|---|---|---|
| 1 | $\gamma=\{0.35,-1.5,0,0.25\}$ $\alpha=\{-0.6,0.41,0.15,0.25\}$ | Linear ($\sigma^2=3$) $\beta=\{0,1,0.5\}$ | IRR: 1.51 | IRR: 1.16 |
| 2 | $\gamma=\{0.35,-1.5,0.25,0.25\}$ $\alpha=\{-0.6,0.41,0.15,0.25\}$ | Linear ($\sigma^2=3$) $\beta=\{0,1,0.5\}$ | IRR: 1.51 | IRR: 1.16 |

*Scenario 1 has a high amount of zero-inflation attributable to the exposure. Scenario 2 keeps the same parameters but increases the amount of zero-inflation attributable to the mediator ($\gamma_2$).

**Table 2: Scenario 1 Simulation Results on a ratio scale***

| Sample Size | Outcome Model | NDE | | | | NIE | | | |
|---|---|---|---|---|---|---|---|---|---|
| | | Bias | Coverage | Power | MSE | Bias | Coverage | Power | MSE |
| **200** | **MZIP** | 1.04 | | | 0.22 | -5.04 | | | 0.07 |
| | Delta | | 94.9% | 46.5% | 0.22 | | 93.8% | 54.5% | 0.07 |
| | Bootstrap | | 94.7% | 49.7% | 0.23 | | 94.6% | 64.6% | 0.07 |
| | **Poisson** | 24.07 | | | 0.53 | -14.11 | | | 0.10 |
| | Delta | | 37.0% | 82.9% | 0.11 | | 77.1% | 57.5% | 0.06 |
| | Bootstrap | | 90.4% | 37.1% | 0.37 | | 93.9% | 47.4% | 0.09 |
| **600** | **MZIP** | 0.61 | | | 0.13 | -1.62 | | | 0.04 |
| | Delta | | 95.5% | 91.6% | 0.13 | | 95.8% | 97.9% | 0.04 |
| | Bootstrap | | 94.3% | 92.4% | 0.13 | | 94.6% | 98.7% | 0.04 |
| | **Poisson** | 20.79 | | | 0.38 | -5.7 | | | 0.07 |
| | Delta | | 27.2% | 92.0% | 0.06 | | 75.2% | 97.3% | 0.04 |
| | Bootstrap | | 88.9% | 51.7% | 0.25 | | 92.7% | 89.9% | 0.05 |
| **1000** | **MZIP** | 0.1 | | | 0.1 | 0.41 | | | 0.03 |
| | Delta | | 94.5% | 99.1% | 0.1 | | 95.0% | 100.0% | 0.03 |
| | Bootstrap | | 94.9% | 99.2% | 0.1 | | 95.2% | 100.0% | 0.03 |
| | **Poisson** | 16.52 | | | 0.37 | -5.78 | | | 0.07 |
| | Delta | | 23.6% | 94.2% | 0.05 | | 70.1% | 99.5% | 0.03 |
| | Bootstrap | | 89.1% | 57.1% | 0.22 | | 93.5% | 83.4% | 0.04 |

*Bias is percent median bias computed as (true value-median estimated)/true value, MSE is the median standard error with the intrinsic value being computed as the standard error between simulated estimates.

**Table 3: Scenario 2 simulation results on a ratio scale***

| Sample Size | Outcome Model | NDE | | | | NIE | | | |
|---|---|---|---|---|---|---|---|---|---|
| | | Bias | Coverage | Power | MSE | Bias | Coverage | Power | MSE |
| 200 | **MZIP** | 1.63 | | | 0.26 | -6.05 | | | 0.08 |
| | Delta | | 94.4% | 36.3% | 0.26 | | 92.4% | 43.2% | 0.08 |
| | Bootstrap | | 94.8% | 39.7% | 0.26 | | 95.0% | 58.9% | 0.08 |
| | **Poisson** | 43.59 | | | 0.76 | -35.78 | | | 0.14 |
| | Delta | | 29.0% | 83.4% | 0.12 | | 59.5% | 48.2% | 0.05 |
| | Bootstrap | | 87.1% | 34.9% | 0.47 | | 84.2% | 18.8% | 0.09 |
| 600 | **MZIP** | 0.87 | | | 0.15 | -2.57 | | | 0.05 |
| | Delta | | 95.4% | 83.8% | 0.15 | | 94.5% | 97.7% | 0.04 |
| | Bootstrap | | 94.4% | 82.8% | 0.15 | | 94.6% | 98.6% | 0.04 |
| | **Poisson** | 30.88 | | | 0.61 | -22.8 | | | 0.11 |
| | Delta | | 21.2% | 91.6% | 0.07 | | 53.9% | 94.3% | 0.03 |
| | Bootstrap | | 86.0% | 46.1% | 0.33 | | 84.0% | 68.3% | 0.06 |
| 1000 | **MZIP** | -0.08 | | | 0.11 | -1.02 | | | 0.03 |
| | Delta | | 95.8% | 96.4% | 0.11 | | 94.2% | 100.0% | 0.03 |
| | Bootstrap | | 94.6% | 96.9% | 0.11 | | 93.8% | 100.0% | 0.03 |
| | **Poisson** | 28.46 | | | 0.55 | -19.23 | | | 0.09 |
| | Delta | | 17.1% | 93.9% | 0.05 | | 51.2% | 98.4% | 0.03 |
| | Bootstrap | | 87.0% | 50.2% | 0.29 | | 84.1% | 80.0% | 0.05 |

*Bias is percent median bias computed as (true value-median estimated)/true value, MSE is the median standard error with the intrinsic value being computed as the standard error between simulated estimates.

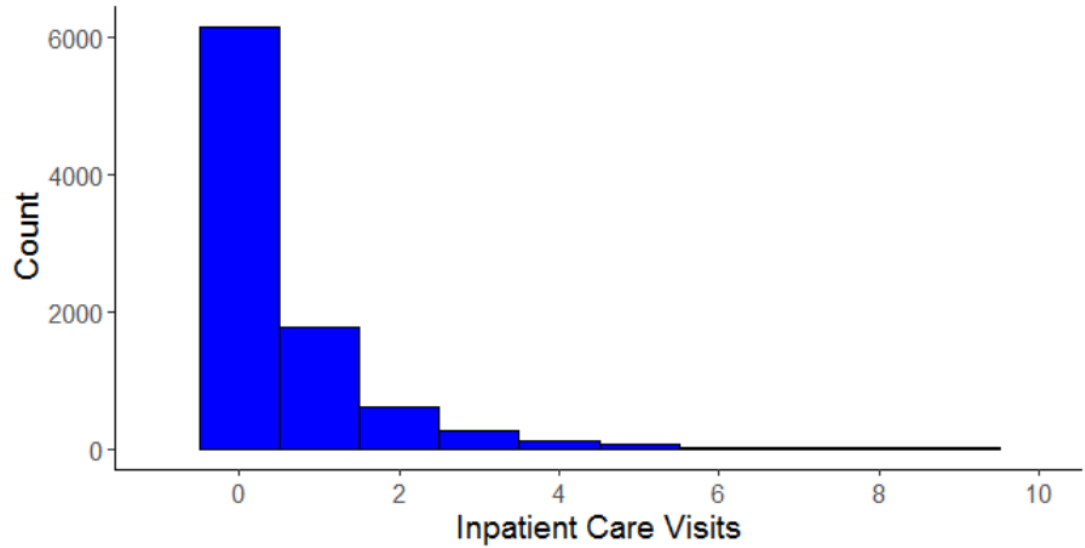

*Figure 2: Distribution of inpatient care visits two years after baseline using REGARDS (2003-2007) Medicare-linked data. The sample size is 9726, about 60% of individuals have zero inpatient care visits.*

**Table 4: Results showing how baseline BMI differences in the number of inpatient care visits 2 years after baseline can be explained by baseline diabetes status using REGARDS (2003-2007).**

| Outcome Model | Estimate | 18.5-<25 kg/m² (n=2613) | 25-<30 kg/m² (n=3729) | 30-<35 kg/m² (n=2047) | 35-<40 kg/m² (n=775) | ≥40 kg/m² (n=484) |
|---|---|---|---|---|---|---|
| **MZIP*** (model) | IRRNIE | Ref | 1.05 (1.04,1.07) | 1.10 (1.08,1.13) | 1.16 (1.12,1.19) | 1.19 (1.14,1.23) |
| | IRRNDE | Ref | 1.02 (0.93,1.11) | 1.10 (0.99,1.22) | 1.07 (0.93,1.24) | 1.23 (1.03,1.46) |
| | IRRTE | Ref | 1.07 (0.97,1.17) | 1.22 (1.10,1.35) | 1.24 (1.07,1.44) | 1.46 (1.23,1.73) |
| | PM | Ref | 76% | 53% | 69% | 50% |
| **MZIP*** (robust) | IRRNIE | Ref | 1.05 (1.03,1.07) | 1.10 (1.08,1.13) | 1.16 (1.12,1.20) | 1.19 (1.14,1.24) |
| | IRRNDE | Ref | 1.02 (0.91,1.13) | 1.10 (0.98,1.24) | 1.07 (0.91,1.27) | 1.23 (1.00,1.52) |
| | IRRTE | Ref | 1.07 (0.96,1.18) | 1.22 (1.08,1.36) | 1.24 (1.05,1.46) | 1.46 (1.18,1.80) |
| | PM | Ref | 76% | 53% | 69% | 50% |
| **Poisson*** | IRRNIE | Ref | 1.05 (1.04,1.07) | 1.11 (1.08,1.14) | 1.16 (1.13,1.19) | 1.19 (1.15,1.23) |
| | IRRNDE | Ref | 1.01 (0.94,1.09) | 1.08 (1.00,1.17) | 1.06 (0.95,1.19) | 1.21 (1.06,1.38) |
| | IRRTE | Ref | 1.06 (0.99,1.14) | 1.20 (1.10,1.30) | 1.23 (1.10,1.38) | 1.44 (1.26,1.64) |
| | PM | Ref | 82% | 59% | 74% | 53% |

*Standard errors computed using delta method

**Supplemental Information Table of Contents**

Web Appendix A: Derivation of Mediation Effects: ZI Outcome, Continuous Mediator

Assume an MZIP outcome and linear regression mediator model with an exposure-mediator interaction.

$$\log(v_i|x, m, \boldsymbol{c}) = \alpha_0 + \alpha_1 x + \alpha_2 m + \alpha_3 xm + \boldsymbol{\alpha_4}'\boldsymbol{c}$$

$$logit(\psi_i|x, m, \boldsymbol{c}) = \gamma_0 + \gamma_1 x + \gamma_2 m + \gamma_3 xm + \boldsymbol{\gamma_4}'\boldsymbol{c}$$

$$E(M_i|x, \boldsymbol{c}) = \theta_0 + \theta_1 x + \boldsymbol{\theta_4}'\boldsymbol{c}$$

We start by deriving the log incidence rate ratio of the NDE

$$\log(IRR^{NDE}) = \sum\{E[Y|x, m, c] - E[Y|x *, m, c]\}P(m|x *, c)$$

$$= \log[P(Y_{XM_{x*}} = y|c)] - \log[P(Y_{X*M_{x*}} = y|c)]$$

Breaking into components we get

$$\log[P(Y_{XM_{x*}} = y|c)] = \log[\int P(Y_{XM} = y|c, M_{x*} = m)P(M_{x*} = m|c)dm]$$

$$= \log[\int P(Y_{XM} = y|c)P(M_{x*} = m|c)dm]$$

$$= \log[\int P(Y = y|x, m, c)P(M = m|x *, c)dm]$$

$$= \log\left[\int e^{(\alpha_0+\alpha_1 X+\alpha_2 M+\alpha_3 XM+\alpha_4 C)} P(M=m|x*,c)dm\right]$$

$$= \log\left[e^{\alpha_0+\alpha_1 x+\alpha_4 C} \int e^{(\alpha_2+\alpha_3 x)m} P(M=m|x*,c)dm\right]$$

$$= \log\left[e^{\alpha_0+\alpha_1 x+\alpha_4 C} E\left(e^{(\alpha_2+\alpha_3 x)m} \big| x*,c\right)\right]$$

Using the moment generating function of the normal distribution we can obtain

$$E\left(e^{(\alpha_2+\alpha_3 x)m} \big| x*,c\right) = e^{\mu_M(\alpha_2+\alpha_3 x)+\frac{\sigma^2(\alpha_2+\alpha_3 x)^2}{2}} = e^{(\alpha_2+\alpha_3 x)(\theta_0+\theta_1 X*+\theta_4 C)+\frac{1}{2}(\alpha_2+\alpha_3 x)^2 \sigma^2}$$

So,

$$\log\left[e^{\alpha_0+\alpha_1 x+\alpha_4 C} E\left(e^{(\alpha_2+\alpha_3 x)m} \big| x*,c\right)\right] = \log\left[e^{\alpha_0+\alpha_1 x+\alpha_4 C} e^{(\alpha_2+\alpha_3 x)(\theta_0+\theta_1 X+\theta_4 C)+\frac{1}{2}(\alpha_2+\alpha_3 x)^2 \sigma^2}\right]$$

$$= \alpha_0 + \alpha_1 x + \alpha_4 C + (\alpha_2+\alpha_3 x)(\theta_0+\theta_1 X*+\theta_4 C) + \frac{1}{2}(\alpha_2+\alpha_3 x)^2 \sigma^2$$

Likewise,

$$P(Y_{XM_x} = y|c) = \alpha_0 + \alpha_1 x + \alpha_4 C + (\alpha_2+\alpha_3 x)(\theta_0+\theta_1 X+\theta_4 C) + \frac{1}{2}(\alpha_2+\alpha_3 x)^2 \sigma^2$$

Therefore,

$$\log(IRR^{NDE}) = \log[P(Y_{XM_{x*}} = y|c)] - \log[P(Y_{X*M_{x*}} = y|c)]$$

$$= \alpha_0 + \alpha_1 x + \alpha_4 C + (\alpha_2+\alpha_3 x)(\theta_0+\theta_1 X*+\theta_4 C) + \frac{1}{2}(\alpha_2+\alpha_3 x)^2 \sigma^2 - [\alpha_0$$

$$+ \alpha_1 x* + \alpha_4 C + (\alpha_2+\alpha_3 x*)(\theta_0+\theta_1 X*+\theta_4 C) + \frac{1}{2}(\alpha_2+\alpha_3 x*)^2 \sigma^2]$$

$$= [\alpha_1 + \alpha_3(\theta_0+\theta_1 x*+\theta_4 c+\alpha_2 \sigma^2)](x-x*) + \frac{1}{2}\alpha_3^2 \sigma^2(x^2 - x*^2)$$

Using the same manipulation for the IRR of the NIE we obtain

$$\log(IRR^{NIE}) = \log[P(Y_{XM_x} = y|c)] - \log[P(Y_{XM_{x*}} = y|c)]$$

$$= \alpha_0 + \alpha_1 x + \alpha_4 C + (\alpha_2 + \alpha_3 x)(\theta_0 + \theta_1 X + \theta_4 C) + \frac{1}{2}(\alpha_2 + \alpha_3 x)^2 \sigma^2$$

$$- \left[\alpha_0 + \alpha_1 x + \alpha_4 C + (\alpha_2 + \alpha_3 x)(\theta_0 + \theta_1 X * + \theta_4 C) + \frac{1}{2}(\alpha_2 + \alpha_3 x)^2 \sigma^2\right]$$

$$= (\alpha_2 \theta_1 + \alpha_3 \theta_1 x)(x - x*)$$

The incidence rate ratio of the CDE can be computed directly

$$IRR^{CDE} = \frac{v_{i,x,m,c}}{v_{i,x*,m,c}} = \frac{e^{\alpha_0 + \alpha_1 X + \alpha_2 M + \alpha_3 XM + \alpha_4 C}}{e^{\alpha_0 + \alpha_1 X* + \alpha_2 M + \alpha_3 X*M + \alpha_4 C}} = \frac{e^{\alpha_1 x + \alpha_3 xm}}{e^{\alpha_1 x* + \alpha_3 x*m}} = e^{(\alpha_1 + \alpha_3 m)(x - x*)}$$

If there is no exposure mediator interaction $\alpha_3$ can be set to zero and derivations will simplify.

Web Appendix B: Derivation of Mediation Effects: ZI Outcome, Continuous Mediator

(Difference Scale)

Assuming the same model expression as in Section A1 without an interaction. NIE and NDE can be expressed on a risk difference scale by.

$$RD(DE) = \sum\{E[Y|x, m, c] - E[Y|x*, m, c]\}P(m|x*, c)$$

$$= \left[e^{\alpha_0 + \alpha_1 x + \alpha_2 m + \alpha_4' C} - e^{\alpha_0 + \alpha_1 x* + \alpha_2 m + \alpha_4' C}\right]P(M_{x*}|C)$$

$$E(e^{\alpha_2 m}) = e^{(\theta_0 + \theta_1 x* + \theta_4' c)\alpha_2 + \frac{\sigma^2 \alpha_2^2}{2}}$$

$$RD(DE) = e^{\alpha_0 + \alpha_4' c + (\theta_0 + \theta_1 x* + \theta_4' c)\alpha_2 + \frac{\sigma^2 \alpha_2^2}{2}}(e^{\alpha_1 x} - e^{\alpha_1 x*})$$

$$RD(IE) = \sum E(Y|x,m)\big(P(m|x) - P(m|x*)\big) = e^{\alpha_0+\alpha_1 x+\alpha_2 m+\alpha_4' C}\big(P(m|x) - P(m|x*)\big)$$

$$= e^{\alpha_0+\alpha_1 x+(\theta_0+\theta_1 x+\theta_4' c)\alpha_2+\frac{\sigma^2 \alpha_2^2}{2}+\alpha_4' C} - e^{\alpha_0+\alpha_1 x+(\theta_0+\theta_1 x*+\theta_4' c)\alpha_2+\frac{\sigma^2 \alpha_2^2}{2}+\alpha_4' C}$$

$$= e^{\alpha_0+\alpha_1 x+(\theta_0+\theta_4' c)\alpha_2+\frac{\sigma^2 \alpha_2^2}{2}+\alpha_4' C}\big(e^{\theta_1 x \alpha_2} - e^{\theta_1 x^* \alpha_2}\big)$$

With an interaction term these derivations are expressed as

$$E(e^{\alpha_2 m+\alpha_3 xm}|m_{x*}) = e^{(\alpha_2+\alpha_3 x)(\theta_0+\theta_1 x*+\theta_4' c)+\frac{\sigma^2(\alpha_2+\alpha_3 x)^2}{2}}$$

$$RD(DE) = e^{\alpha_0+\alpha_1 x+(\alpha_2+\alpha_3 x)(\theta_0+\theta_1 x*+\theta_4' c)+\frac{\sigma^2(\alpha_2+\alpha_3 x)^2}{2}+\alpha_4' C}$$

$$- e^{\alpha_0+\alpha_1 x*+(\alpha_2+\alpha_3 x*)(\theta_0+\theta_1 x*+\theta_4' c)+\frac{\sigma^2(\alpha_2+\alpha_3 x*)^2}{2}+\alpha_4' C}$$

$$RD(IE) = e^{\alpha_0+\alpha_1 x+(\alpha_2+\alpha_3 x)(\theta_0+\theta_1 x+\theta_4' c)+\frac{\sigma^2(\alpha_2+\alpha_3 x)^2}{2}+\alpha_4' C}$$

$$- e^{\alpha_0+\alpha_1 x+(\alpha_2+\alpha_3 x)(\theta_0+\theta_1 x*+\theta_4' c)+\frac{\sigma^2(\alpha_2+\alpha_3 x)^2}{2}+\alpha_4' C}$$

Web Appendix C: Delta Method Standard Error Derivations for ZI Outcome, Continuous Mediator

*Ratio Scale*

In this section we will show derivations for the delta method standard errors for the natural direct and natural indirect effect in the previous section. When the outcome is a zero-inflated count and direct and indirect effects are on a risk difference scale it is easier to derive standard errors using the log-IRR of the effects. We will the following formula for standard error estimation

$$\sqrt{\Gamma \sum \Gamma} |x - x*|$$

Where,

$$\Sigma = \begin{vmatrix} \Sigma_\theta & 0 & 0 \\ 0 & \Sigma_\alpha & 0 \\ 0 & 0 & \Sigma_{\sigma^2} \end{vmatrix}$$

where $\Sigma_\theta$ and $\Sigma_\alpha$ are the covariance matrices for the estimators of $\theta$ and $\alpha$. $\Gamma$ is a vector of partial derivatives of the formula for each effect with respect to each parameter in the model. In this case we have the extra term of $\Sigma_{\sigma^2}$ because some of the effects depends upon $\sigma^2$, the variance of the linear model. $\Sigma_{\sigma^2}$ is the covariance matrix for the estimator of $\sigma^2$. We previously noted a formula for an unbiased estimate of $\sigma^2$. The variance of $\sigma^2$ can be estimated by

$$\frac{2\hat{\sigma}^4}{n-p}$$

where $n$ is the sample size and $p$ is the number of parameters in the model.

Next, we need to derive $\Gamma$ for each of the effects by taking the partial derivative with respect to $\theta, \alpha,$ and $\sigma^2$ coefficient in the formula for the effect. We start with the controlled direct effect.

$$\Gamma(\log(IRR^{CDE})) = (0,0,0',0,1,0,m,0',0)$$

Where $0'$ is a row vector zeroes of length equal to the length of $C$.

For the natural direct effect we obtain:

$$\begin{aligned}\Gamma(\log(IRR^{NDE})) \\ = \big(\alpha_3, \alpha_3 x*, \alpha_3' c, 0, 1, \alpha_3 \sigma^2, \theta_0 + \theta_1 x* + \theta_2' C + \alpha_2 \sigma^2 \\ + \alpha_2 \sigma^2(x + x*), 0', 0.5\alpha_3^2(x + x*)\big)\end{aligned}$$

For the natural indirect effect we have:

$$\Gamma(\log(IRR^{NIE})) = (0, \alpha_2 + \alpha_3 x, 0', 0, 0, \theta_1, \theta_1 x, 0', 0)$$

The $\Gamma$ for the total effect can be computed as the sum of the $\Gamma$'s for the log of the natural direct and natural indirect effect.

$$\Gamma(\log(IRR^{TE})) = (t1, t2, t3, t4, t5, t6, t7, t8, t9)$$

$$t1 = \alpha_3$$

$$t2 = (\alpha_2 + \alpha_3(x + x*))$$

$$t3 = \alpha_3 C$$

$$t4 = 0$$

$$t5 = 1$$

$$t6 = \theta_1$$

$$t7 = (\theta_{0\_} + \theta_1(x + x*) + \theta_4' C \sigma^2) + \alpha_3 \sigma^2 (x + x*)$$

$$t8 = 0$$

$$t9 = \alpha_3 \theta_4' C + \frac{1}{2}\alpha_3^2 (x + x*)$$

Note: If there is no exposure-mediator interaction then CDE and NDE will be equivalent. NDE and NIE will simplify to

$$IRR^{NDE} = e^{\alpha_1(x-x*)}$$

$$IRR^{NIE} = e^{\alpha_2 \theta_1 (x-x*)}$$

Since these formulas are not dependent on $\sigma^2$, we will use

$$\Sigma = \begin{bmatrix} \Sigma_\theta & 0 \\ 0 & \Sigma_\alpha \end{bmatrix}$$

For our delta method derivations. In this case

$$\Gamma(\log(IRR^{NDE})) = (0,0,0', 0,1,0,0,0')$$

$$\Gamma(\log(IRR^{NIE})) = (0, \alpha_2, 0', 0, 0, \theta_1, 0')$$

Where 0' is a row vector of the dimensions of C, filled with zeroes.

To compute 95% confidence intervals for the direct and indirect effect based of the calculated standard errors of the log incidence rate ratio we use the following equation

$$e^{\log(IRR^{NDE}) \pm 1.96\sqrt{SE(\log(IRR^{NDE}))}}$$

Where SE is the standard error of the respective mediation effect's standard error. We used a 95% confidence interval, but other levels of significance can be used in the place of 1.96.

If the confidence interval computed includes 1 in the interval, then the effect is not statistically significant; otherwise, the effect is statistically significant.

*Difference Scale*

On a difference scale we will the following formula for standard error estimation

$$\sqrt{\Gamma \Sigma \Gamma}$$

where everything is the same as it was on a ratio scale except

$$\Gamma(RD(DE)) = (d1, d2, d3, d4, d5, d6, d7, d8, d9)$$

$$d1 = (\alpha_2 + \alpha_3 x)e^{\alpha_0+\alpha_1 x+(\alpha_2+\alpha_3 x)(\theta_0+\theta_1 x*+\theta_4'c)+\frac{\sigma^2(\alpha_2+\alpha_3 x)^2}{2}+\alpha_4'C} - (\alpha_2 + \alpha_3 x^*)e^{\alpha_0+\alpha_1 x*+(\alpha_2+\alpha_3 x*)(\theta_0+\theta_1 x*+\theta_4'c)+\frac{\sigma^2(\alpha_2+\alpha_3 x*)^2}{2}+\alpha_4'C}$$

$$d2 = x^* d1$$

$$d3 = c'd1$$

$$d4 = RD(DE)$$

$$d5 = xe^{\alpha_0+\alpha_1 x+(\alpha_2+\alpha_3 x)(\theta_0+\theta_1 x*+\theta_4'c)+\frac{\sigma^2(\alpha_2+\alpha_3 x)^2}{2}+\alpha_4'C} - x^* e^{\alpha_0+\alpha_1 x*+(\alpha_2+\alpha_3 x*)(\theta_0+\theta_1 x*+\theta_4'c)+\frac{\sigma^2(\alpha_2+\alpha_3 x*)^2}{2}+\alpha_4'C}$$

$$d6 = \left(\theta_0 + \theta_1 x^* + \theta_4'c + \sigma^2(\alpha_2 + \alpha_3 x)\right)e^{\alpha_0+\alpha_1 x+(\alpha_2+\alpha_3 x)(\theta_0+\theta_1 x*+\theta_4'c)+\frac{\sigma^2(\alpha_2+\alpha_3 x)^2}{2}+\alpha_4'C} - \left(\theta_0 + \theta_1 x^* + \theta_4'c + \sigma^2(\alpha_2 + \alpha_3 x^*)\right)e^{\alpha_0+\alpha_1 x*+(\alpha_2+\alpha_3 x*)(\theta_0+\theta_1 x*+\theta_4'c)+\frac{\sigma^2(\alpha_2+\alpha_3 x*)^2}{2}+\alpha_4'C}$$

$$d7 = x\left(\theta_0 + \theta_1 x^* + \theta_4'c + \sigma^2(\alpha_2 + \alpha_3 x)\right)e^{\alpha_0+\alpha_1 x+(\alpha_2+\alpha_3 x)(\theta_0+\theta_1 x*+\theta_4'c)+\frac{\sigma^2(\alpha_2+\alpha_3 x)^2}{2}+\alpha_4'C} - x^*\left(\theta_0 + \theta_1 x^* + \theta_4'c + \sigma^2(\alpha_2 + \alpha_3 x^*)\right)e^{\alpha_0+\alpha_1 x*+(\alpha_2+\alpha_3 x*)(\theta_0+\theta_1 x*+\theta_4'c)+\frac{\sigma^2(\alpha_2+\alpha_3 x*)^2}{2}+\alpha_4'C}$$

$$d8 = c'RD(DE)$$

$$d9 = \frac{(\alpha_2 + \alpha_3 x)^2}{2}e^{\alpha_0+\alpha_1 x+(\alpha_2+\alpha_3 x)(\theta_0+\theta_1 x*+\theta_4'c)+\frac{\sigma^2(\alpha_2+\alpha_3 x)^2}{2}+\alpha_4'C} - \frac{(\alpha_2 + \alpha_3 x^*)^2}{2}e^{\alpha_0+\alpha_1 x*+(\alpha_2+\alpha_3 x*)(\theta_0+\theta_1 x*+\theta_4'c)+\frac{\sigma^2(\alpha_2+\alpha_3 x*)^2}{2}+\alpha_4'C}$$

$$\Gamma(RD(IE)) = (i1, i2, i3, i4, i5, i6, i7, i8, i9)$$

$$i1 = (\alpha_2 + \alpha_3 x)RD(IE)$$

$$i2 = x(\alpha_2 + \alpha_3 x)e^{\alpha_0+\alpha_1 x+(\alpha_2+\alpha_3 x)(\theta_0+\theta_1 x+\theta'_4 c)+\frac{\sigma^2(\alpha_2+\alpha_3 x)^2}{2}+\alpha'_4 c}$$

$$- x^*(\alpha_2 + \alpha_3 x)e^{\alpha_0+\alpha_1 x+(\alpha_2+\alpha_3 x)(\theta_0+\theta_1 x*+\theta'_4 c)+\frac{\sigma^2(\alpha_2+\alpha_3 x)^2}{2}+\alpha'_4 c}$$

$$i3 = c'i1$$

$$i4 = RD(IE)$$

$$i5 = xRD(IE)$$

$$i6 = \left(\theta_0 + \theta_1 x + \theta'_4 c + \sigma^2(\alpha_2 + \alpha_3 x)\right)e^{\alpha_0+\alpha_1 x+(\alpha_2+\alpha_3 x)(\theta_0+\theta_1 x+\theta'_4 c)+\frac{\sigma^2(\alpha_2+\alpha_3 x)^2}{2}+\alpha'_4 c}$$

$$- \left(\theta_0 + \theta_1 x^* + \theta'_4 c\right.$$

$$\left.+ \sigma^2(\alpha_2 + \alpha_3 x)\right)e^{\alpha_0+\alpha_1 x+(\alpha_2+\alpha_3 x)(\theta_0+\theta_1 x*+\theta'_4 c)+\frac{\sigma^2(\alpha_2+\alpha_3 x)^2}{2}+\alpha'_4 c}$$

$$i7 = xi6$$

$$i8 = c'RD(IE)$$

$$i9 = \frac{(\alpha_2 + \alpha_3 x)^2}{2}RD(IE)$$

To compute 95% confidence intervals for the direct and indirect effect based of the calculated standard errors of the log incidence rate ratio, we use the following equation

$$RD^{NDE} \pm 1.96\sqrt{SE(RD^{NDE})}$$

Web Appendix D: Derivation of Mediation Effects: ZI Outcome, Binary Mediator

For a binary mediator, assuming an exposure-mediator interaction, we specify the following models with an MZIP outcome and logistic regression mediator model

$$logit(\psi|x, m, \boldsymbol{c}) = \gamma_0 + \gamma_1 x + \gamma_2 m + \gamma_3 xm + \boldsymbol{\gamma_4'C}$$

$$\log(\nu|x, m, \boldsymbol{c}) = \alpha_0 + \alpha_1 x + \alpha_2 m + \alpha_3 xm + \boldsymbol{\alpha_4'C}$$

$$logit\{P(M = 1|x, \boldsymbol{c})\} = \theta_0 + \theta_1 x + \boldsymbol{\theta_4'C}$$

Effects are specified as

$$IRR^{NDE} = \frac{\left(e^{\alpha_1 x}\left(1 + e^{\alpha_2 + \alpha_3 x + \theta_0 + \theta_1 x* + \theta_2' C}\right)\right)}{\left(e^{\alpha_1 x*}\left(1 + e^{\alpha_2 + \alpha_3 a* + \theta_0 + \theta_1 x* + \theta_2' C}\right)\right)}$$

$$IRR^{NIE} = \frac{\left([1 + e^{\theta_0 + \theta_1 x* + \theta_2' C}][1 + e^{\alpha_2 + \alpha_3 x + \theta_0 + \theta_1 x + \theta_2' C}]\right)}{\left([1 + e^{\theta_0 + \theta_1 x + \theta_2' C}][1 + e^{\alpha_2 + \alpha_3 x + \theta_0 + \theta_1 x* + \theta_2' C}]\right)}$$

Proofs of these effects are demonstrated below. The controlled direct effect is computed the same way as in the case of a continuous mediator. For the other effects we need to use expressions similar to

What is $\log[P(Y_{XM_{x*}} = y|c)]$?

$$\log[P(Y_{XM_{x*}} = y|c)] = \log[\int P(Y_{XM} = y|c, M_{x*} = m)P(M_{x*} = m|c)dm]$$

$$= \log[\int P(Y_{XM} = y \,|c)P(M_{x*} = m|c)dm]$$

$$= \log [\int P(Y = y|x, m, c)P(M = m|x *, c)dm]$$

$$= \log\left[\int e^{(\alpha_0+\alpha_1 X+\alpha_2 M+\alpha_3 XM+\alpha_4 C)} P(M=m|x*,c)dm\right]$$

$$= \log\left[e^{\alpha_0+\alpha_1 x+\alpha_4 C}\int e^{(\alpha_2+\alpha_3 x)m} P(M=m|x*,c)dm\right]$$

$$= \log\left[e^{\alpha_0+\alpha_1 x+\alpha_4 C} E\left(e^{(\alpha_2+\alpha_3 x)m}\big|x*,c\right)\right]$$

Since the mediator follows a Bernoulli distribution with a logit-link

$$E\left(e^{(\alpha_2+\alpha_3 x)m}\big|x*,c\right) = \left(1 - \frac{e^{\theta_0+\theta_1 x*+\theta_2' c}}{1+e^{\theta_0+\theta_1 x*+\theta_2' c}} + \frac{e^{\theta_0+\theta_1 x*+\theta_2' c}}{1+e^{\theta_0+\theta_1 x*+\theta_2' c}} e^{\alpha_2+\alpha_3 x}\right)$$

$$= \left(\left(\frac{1}{1+e^{\theta_0+\theta_1 x*+\theta_2' c}}\right) + \frac{e^{\theta_0+\theta_1 x*+\theta_2' c}}{1+e^{\theta_0+\theta_1 x*+\theta_2' c}} e^{\alpha_2+\alpha_3 x}\right)$$

So,

$$\log\left[P(Y_{XM_{x*}} = y|c)\right] = \log\left[e^{\alpha_0+\alpha_1 x+\alpha_4 C'}\left(\left(\frac{1}{1+e^{\theta_0+\theta_1 x*+\theta_2' c}}\right) + \frac{e^{\theta_0+\theta_1 x*+\theta_2' c}}{1+e^{\theta_0+\theta_1 x*+\theta_2' c}} e^{\alpha_2+\alpha_3 x}\right)\right]$$

For the natural direct effect we have

$$IRR^{(NDE)} = e^{\log\left[P(Y_{XM_{x*}}=y|c)\right]-\log\left[P(Y_{X*M_{x*}}=y|c)\right]}$$

$$= \frac{e^{\alpha_0+\alpha_1 x+\alpha_4 C'}\left(\left(\frac{1}{1+e^{\theta_0+\theta_1 x*+\theta_2' c}}\right) + \frac{e^{\theta_0+\theta_1 x*+\theta_2' c}}{1+e^{\theta_0+\theta_1 x*+\theta_2' c}} e^{\alpha_2+\alpha_3 x}\right)}{e^{\alpha_0+\alpha_1 x*+\alpha_4 C'}\left(\left(\frac{1}{1+e^{\theta_0+\theta_1 x*+\theta_2' c}}\right) + \frac{e^{\theta_0+\theta_1 x*+\theta_2' c}}{1+e^{\theta_0+\theta_1 x*+\theta_2' c}} e^{\alpha_2+\alpha_3 x*}\right)}$$

$$= \frac{\left(e^{\alpha_1 x}\left(1+e^{\alpha_2+\alpha_3 x+\theta_0+\theta_1 x*+\theta_2' C}\right)\right)}{\left(e^{\alpha_1 x*}\left(1+e^{\alpha_2+\alpha_3 a*+\theta_0+\theta_1 x*+\theta_2' C}\right)\right)}$$

For the natural indirect effect, we have

$$IRR^{(NIE)} = e^{\log[P(Y_{XM_x} = y|c)] - \log[P(Y_{XM_{x*}} = y|c)]}$$

$$= \frac{e^{\alpha_0 + \alpha_1 x + \alpha_4 C'} \left( \left( \frac{1}{1 + e^{\theta_0 + \theta_1 x + \theta_2' c}} \right) + \frac{e^{\theta_0 + \theta_1 x + \theta_2' c}}{1 + e^{\theta_0 + \theta_1 x + \theta_2' c}} e^{\alpha_2 + \alpha_3 x} \right)}{e^{\alpha_0 + \alpha_1 x + \alpha_4 C'} \left( \left( \frac{1}{1 + e^{\theta_0 + \theta_1 x* + \theta_2' c}} \right) + \frac{e^{\theta_0 + \theta_1 x* + \theta_2' c}}{1 + e^{\theta_0 + \theta_1 x* + \theta_2' c}} e^{\alpha_2 + \alpha_3 x} \right)}$$

$$= \frac{\left([1 + e^{\theta_0 + \theta_1 x* + \theta_2' C}][1 + e^{\alpha_2 + \alpha_3 x + \theta_0 + \theta_1 x + \theta_2' C}]\right)}{\left([1 + e^{\theta_0 + \theta_1 x + \theta_2' C}][1 + e^{\alpha_2 + \alpha_3 x + \theta_0 + \theta_1 x* + \theta_2' C}]\right)}$$

Web Appendix E: Derivation of Effects: ZI Outcome, Binary Mediator on a Difference Scale

Using the same model expression as in A4 and definitions in A2 the following expressions can be derived for mediation effects with a MZIP outcome model and logistic mediator on a difference scale.

$$E(e^{\alpha_2 m + \alpha_3 xm}) = \frac{1}{1 + e^{\theta_0 + \theta_1 x* + \theta_4' c}} + \frac{e^{\theta_0 + \theta_1 x* + \theta_4' c}}{1 + e^{\theta_0 + \theta_1 x* + \theta_4' c}} e^{\alpha_2 + \alpha_3 x}$$

$$RD(DE) = e^{\alpha_0 + \alpha_1 x + \alpha_4' C} \left( \frac{1}{1 + e^{\theta_0 + \theta_1 x* + \theta_4' c}} + \frac{e^{\theta_0 + \theta_1 x* + \theta_4' c}}{1 + e^{\theta_0 + \theta_1 x* + \theta_4' c}} e^{\alpha_2 + \alpha_3 x} \right)$$

$$- e^{\alpha_0 + \alpha_1 x* + \alpha_4' C} \left( \frac{1}{1 + e^{\theta_0 + \theta_1 x* + \theta_4' c}} + \frac{e^{\theta_0 + \theta_1 x* + \theta_4' c}}{1 + e^{\theta_0 + \theta_1 x* + \theta_4' c}} e^{\alpha_2 + \alpha_3 x*} \right)$$

$$RD(IE) = e^{\alpha_0 + \alpha_1 x + \alpha_4' C} \left( \frac{1}{1 + e^{\theta_0 + \theta_1 x + \theta_4' c}} + \frac{e^{\theta_0 + \theta_1 x + \theta_4' c}}{1 + e^{\theta_0 + \theta_1 x + \theta_4' c}} e^{\alpha_2 + \alpha_3 x} \right)$$

$$- e^{\alpha_0 + \alpha_1 x + \alpha_4' C} \left( \frac{1}{1 + e^{\theta_0 + \theta_1 x* + \theta_4' c}} + \frac{e^{\theta_0 + \theta_1 x* + \theta_4' c}}{1 + e^{\theta_0 + \theta_1 x* + \theta_4' c}} e^{\alpha_2 + \alpha_3 x} \right)$$

Web Appendix F: Delta Method Standard Error Derivations for ZI Outcome, Binary Mediator

*Ratio Scale*

For the delta method derivations we will use the formula

$$\sqrt{\Gamma \Sigma \Gamma}$$

Where,

$$\Sigma = \begin{bmatrix} \Sigma_\theta & 0 \\ 0 & \Sigma_\alpha \end{bmatrix}$$

Where $\Sigma_\theta$ and $\Sigma_\alpha$ are the covariance matrices for the estimators of $\theta$ and $\alpha$. $\Gamma$ for the natural log of the IRR of each effect can be computed as follows

$$\Gamma(logIRR^{CDE}) = (0,0,0',0,1,0,m,0)(x - x*)$$

For the natural direct effect we first set let

$$Q = \frac{e^{\alpha_2 + \alpha_3 x + \theta_0 + \theta_1 x* + \theta_2' C}}{[1 + e^{\alpha_2 + \alpha_3 x + \theta_0 + \theta_1 x* + \theta_2' C}]}$$

$$B = \frac{e^{\alpha_2 + \alpha_3 x* + \theta_0 + \theta_1 x* + \theta_2' C}}{[1 + e^{\alpha_2 + \alpha_3 x* + \theta_0 + \theta_1 x* + \theta_2' C}]}$$

$$\Gamma(logIRR^{NDE}) = (Q - B, x * (Q - B), C'(Q - B), 0, x - x*, Q - B. xQ - x * B, 0')$$

For the natural indirect effect we let

$$R = \frac{e^{\alpha_2 + \alpha_3 x + \theta_0 + \theta_1 x + \theta_2' C}}{[1 + e^{\alpha_2 + \alpha_3 x + \theta_0 + \theta_1 x + \theta_2' C}]}$$

$$F = \frac{e^{\alpha_2+\alpha_3 x+\theta_0+\theta_1 x*+\theta_2'C}}{\left[1 + e^{\alpha_2+\alpha_3 x+\theta_0+\theta_1 x*+\theta_2'C}\right]}$$

$$K = \frac{e^{\theta_0+\theta_1 x+\theta_2'C}}{1 + e^{\theta_0+\theta_1 x+\theta_2'C}}$$

$$D = \frac{e^{\theta_0+\theta_1 x*+\theta_2'C}}{1 + e^{\theta_0+\theta_1 x*+\theta_2'C}}$$

$$\Gamma(logIRR^{NIE}) = (d1, d2, d3, d4, d5, d6, d7, d8)$$

$$d1 = D + R - K - F$$

$$d2 = x * (D - F) + x(R - K)$$

$$d3 = C'(D + R - K - F)$$

$$d4 = 0$$

$$d5 = 0$$

$$d6 = R - F$$

$$d7 = x(R - F)$$

$$d8 = 0'$$

For the total effect we have

$$\Gamma(logIRR(TE)) = \Gamma(\log IRR^{NDE}) + \Gamma(logIRR^{NIE})$$

*Difference Scale*

On a difference scale everything is the same as on a ratio scale except

$$RD(DE) = e^{\alpha_0+\alpha_1 x+\alpha_4' C}\left(\frac{1}{1+e^{\theta_0+\theta_1 x*+\theta_4' c}}+\frac{e^{\theta_0+\theta_1 x*+\theta_4' c}}{1+e^{\theta_0+\theta_1 x*+\theta_4' c}}e^{\alpha_2+\alpha_3 x}\right)$$

$$-e^{\alpha_0+\alpha_1 x*+\alpha_4' C}\left(\frac{1}{1+e^{\theta_0+\theta_1 x*+\theta_4' c}}+\frac{e^{\theta_0+\theta_1 x*+\theta_4' c}}{1+e^{\theta_0+\theta_1 x*+\theta_4' c}}e^{\alpha_2+\alpha_3 x*}\right)$$

$$\Gamma\big(RD(DE)\big) = (d1, d2, d3, d4, d5, d6, d7, d8)$$

$d1$

$$= \left(e^{\alpha_0+\alpha_1 x+\alpha_4' c}\right)$$

$$* \left(\frac{\left(1+e^{(\theta_0+\theta_1 x*+\theta_4' c)}\right)e^{\theta_0+\theta_1 x*+\theta_4 c'+\alpha_2+\alpha_3 x} - e^{\theta_0+\theta_1 x*+\theta_4 c'}\left(1+e^{\theta_0+\theta_1 x*+\theta_3' c+\alpha_2+\alpha_3 x}\right)}{\left(1+e^{\theta_0+\theta_1 x*+\theta_4' c}\right)^2}\right)$$

$$-\left(e^{\alpha_0+\alpha_1 x*+\alpha_4' c}\right)$$

$$* \left(\frac{\left(1+e^{(\theta_0+\theta_1 x*+\theta_4' c)}\right)e^{\theta_0+\theta_1 x*+\theta_4 c'+\alpha_2+\alpha_3 x*} - e^{\theta_0+\theta_1 x*+\theta_4 c'}\left(1+e^{\theta_0+\theta_1 x*+\theta_3' c+\alpha_2+\alpha_3 x*}\right)}{\left(1+e^{\theta_0+\theta_1 x*+\theta_4' c}\right)^2}\right)$$

$$d2 = x^* d1$$

$$d3 = c' d1$$

$$d4 = RD(DE)$$

$$d5 = xe^{\alpha_0+\alpha_1 x+\alpha_4' C}\left(\frac{1}{1+e^{\theta_0+\theta_1 x*+\theta_4' c}}+\frac{e^{\theta_0+\theta_1 x*+\theta_4' c}}{1+e^{\theta_0+\theta_1 x*+\theta_4' c}}e^{\alpha_2+\alpha_3 x}\right)$$

$$-x^* e^{\alpha_0+\alpha_1 x*+\alpha_4' C}\left(\frac{1}{1+e^{\theta_0+\theta_1 x*+\theta_4' c}}+\frac{e^{\theta_0+\theta_1 x*+\theta_4' c}}{1+e^{\theta_0+\theta_1 x*+\theta_4' c}}e^{\alpha_2+\alpha_3 x*}\right)$$

$$d6 = d1$$

$d7$

$$= \left(xe^{\alpha_0+\alpha_1 x+\alpha'_4 c}\right)$$

$$* \left(\frac{\left(1+e^{(\theta_0+\theta_1 x*+\theta'_4 c)}\right)e^{\theta_0+\theta_1 x^*+\theta_4 c'+\alpha_2+\alpha_3 x} - e^{\theta_0+\theta_1 x^*+\theta_4 c'}\left(1+e^{\theta_0+\theta_1 x^*+\theta'_3 c+\alpha_2+\alpha_3 x}\right)}{\left(1+e^{\theta_0+\theta_1 x*+\theta'_4 c}\right)^2}\right)$$

$$- \left(x^* e^{\alpha_0+\alpha_1 x^*+\alpha'_4 c}\right)$$

$$* \left(\frac{\left(1+e^{(\theta_0+\theta_1 x*+\theta'_4 c)}\right)e^{\theta_0+\theta_1 x^*+\theta_4 c'+\alpha_2+\alpha_3 x^*} - e^{\theta_0+\theta_1 x^*+\theta_4 c'}\left(1+e^{\theta_0+\theta_1 x^*+\theta'_3 c+\alpha_2+\alpha_3 x^*}\right)}{\left(1+e^{\theta_0+\theta_1 x*+\theta'_4 c}\right)^2}\right)$$

$$d8 = c' RD(DE)$$

$$RD(IE) = e^{\alpha_0+\alpha_1 x+\alpha'_4 C}\left(\frac{1}{1+e^{\theta_0+\theta_1 x+\theta'_4 c}} + \frac{e^{\theta_0+\theta_1 x+\theta'_4 c}}{1+e^{\theta_0+\theta_1 x+\theta'_4 c}}e^{\alpha_2+\alpha_3 x}\right)$$

$$- e^{\alpha_0+\alpha_1 x+\alpha'_4 C}\left(\frac{1}{1+e^{\theta_0+\theta_1 x*+\theta'_4 c}} + \frac{e^{\theta_0+\theta_1 x*+\theta'_4 c}}{1+e^{\theta_0+\theta_1 x*+\theta'_4 c}}e^{\alpha_2+\alpha_3 x}\right)$$

$$\Gamma\bigl(RD(IE)\bigr) = (i1, i2, i3, i4, i5, i6, i7, i8)$$

$i1$

$$= \left(e^{\alpha_0+\alpha_1 x+\alpha'_4 C}\right)$$

$$* \left(\frac{\left(1+e^{(\theta_0+\theta_1 x+\theta'_4 c)}\right)e^{\theta_0+\theta_1 x+\theta_4 c'+\alpha_2+\alpha_3 x} - e^{\theta_0+\theta_1 x+\theta_4 c'}\left(1+e^{\theta_0+\theta_1 x+\theta'_3 c+\alpha_2+\alpha_3 x}\right)}{\left(1+e^{\theta_0+\theta_1 x+\theta'_4 c}\right)^2}\right)$$

$$- \left(e^{\alpha_0+\alpha_1 x+\alpha'_4 C}\right)$$

$$* \left(\frac{\left(1+e^{(\theta_0+\theta_1 x*+\theta'_4 c)}\right)e^{\theta_0+\theta_1 x^*+\theta_4 c'+\alpha_2+\alpha_3 x} - e^{\theta_0+\theta_1 x^*+\theta_4 c'}\left(1+e^{\theta_0+\theta_1 x^*+\theta'_3 c+\alpha_2+\alpha_3 x}\right)}{\left(1+e^{\theta_0+\theta_1 x*+\theta'_4 c}\right)^2}\right)$$

$$i2$$

$$= \left(xe^{\alpha_0+\alpha_1 x+\alpha_4' C}\right)$$

$$* \left(\frac{\left(1+e^{(\theta_0+\theta_1 x+\theta_4' c)}\right)e^{\theta_0+\theta_1 x+\theta_4 c'+\alpha_2+\alpha_3 x} - e^{\theta_0+\theta_1 x+\theta_4 c'}\left(1+e^{\theta_0+\theta_1 x+\theta_3' c+\alpha_2+\alpha_3 x}\right)}{\left(1+e^{\theta_0+\theta_1 x+\theta_4' c}\right)^2}\right)$$

$$- \left(x^* e^{\alpha_0+\alpha_1 x+\alpha_4' C}\right)$$

$$* \left(\frac{\left(1+e^{(\theta_0+\theta_1 x*+\theta_4' c)}\right)e^{\theta_0+\theta_1 x^*+\theta_4 c'+\alpha_2+\alpha_3 x} - e^{\theta_0+\theta_1 x^*+\theta_4 c'}\left(1+e^{\theta_0+\theta_1 x^*+\theta_3' c+\alpha_2+\alpha_3 x}\right)}{\left(1+e^{\theta_0+\theta_1 x*+\theta_4' c}\right)^2}\right)$$

$$i3 = c'i1$$

$$i4 = RD(IE)$$

$$i5 = xRD(IE)$$

$$i6 = i1$$

$$i7 = xi1$$

$$i8 = c'i1$$

Web Appendix G: Supplemental Simulations for continuous mediator

Many more scenarios were considered in our simulation study for the proposed method displayed below.

**Web Table 1. Supplemental simulation scenarios**

| Scenario | MZIP Parameters | Mediator Model | NDE | NIE |
|---|---|---|---|---|
| 3 | $\gamma$={0.35,-.45,0,0.25} | Linear ($\sigma^2$=3) | IRR: 1.51 | IRR: 1.16 |
|  | $\alpha$={-0.6,0.41,0.6,0.25} | $\beta$={0,1,0.5} | RD: 0.54 | RD: 0.27 |
| 4 | $\gamma$={0.35,-1.5,0,0.25} | Linear ($\sigma^2$=3) | IRR: 1.51 | IRR: 1.16 |
| | $\alpha$={0.4,0.41,0.6,0.25} | $\beta$={0,1,0.5} | RD: 1.47 | RD: 0.72 |
| 5 | $\gamma$={0.35,-1.5,0.25,0.25} | Linear ($\sigma^2$=3) | IRR: 1.51 | IRR: 1.16 |
| | $\alpha$={0.4,0.41,0.6,0.25} | $\beta$={0,1,0.5} | RD: 1.47 | RD: 0.72 |

Scenarios 3-5 expand upon Scenarios 1 and 2 from Table 1 in the main text. Scenario 3 reduces the impact of the exposure on zero-inflation $\gamma_1$ from Scenario 1. Scenarios 4 and 5 increase the overall population mean, $\alpha_0$, from Scenario 1 and 2. Results from these simulations are depicted in the following sections on both ratio and difference scale. Table S2-S4 show results from Scenarios 3-5 on a ratio scale and Tables S5-S9 depict results from all 5 scenarios on a difference scale.

*Ratio Scale*

**Web Table 2: Scenario 3 simulation results on a ratio scale**

| Sample Size | Outcome Model | NDE | | | | NIE | | | |
|---|---|---|---|---|---|---|---|---|---|
| | | Bias | Coverage | Power | MSE | Bias | Coverage | Power | MSE |
| **200** | **MZIP** | 1.14 | | | 0.25 | -4.2 | | | 0.08 |
| | Delta | | 94.7% | 39.3% | 0.24 | | 93.0% | 46.1% | 0.08 |
| | Bootstrap | | 95.4% | 39.3% | 0.25 | | 94.5% | 60.5% | 0.08 |
| | **Poisson** | 11.09 | | | 0.6 | -16.26 | | | 0.11 |
| | Delta | | 34.1% | 78.4% | 0.11 | | 73.0% | 55.9% | 0.06 |
| | Bootstrap | | 92.3% | 25.7% | 0.40 | | 92.2% | 35.5% | 0.10 |
| **600** | **MZIP** | 0.32 | | | 0.14 | -1.47 | | | 0.04 |
| | Delta | | 94.5% | 83.0% | 0.14 | | 94.3% | 99.0% | 0.04 |
| | Bootstrap | | 94.7% | 86.0% | 0.14 | | 94.1% | 99.0% | 0.04 |
| | **Poisson** | 10.24 | | | 0.46 | -7.64 | | | 0.08 |
| | Delta | | 28.0% | 88.8% | 0.06 | | 68.3% | 97.4% | 0.04 |
| | Bootstrap | | 92.8% | 40.0% | 0.28 | | 92.4% | 84.4% | 0.06 |
| **1000** | **MZIP** | -0.13 | | | 0.11 | 0.53 | | | 0.03 |
| | Delta | | 95.1% | 98.0% | 0.11 | | 94.2% | 100.0% | 0.03 |
| | Bootstrap | | 95.2% | 97.4% | 0.11 | | 96.0% | 100.0% | 0.03 |
| | **Poisson** | 8.56 | | | 0.42 | -4.75 | | | 0.07 |
| | Delta | | 25.0% | 93.6% | 0.05 | | 64.6% | 99.4% | 0.03 |
| | Bootstrap | | 92.2% | 49.1% | 0.25 | | 95.0% | 90.2% | 0.05 |

**Web Table 3: Scenario 4 simulation results on a ratio scale**

| Sample Size | Outcome Model | NDE | | | | NIE | | | |
|---|---|---|---|---|---|---|---|---|---|
| | | Bias | Coverage | Power | MSE | Bias | Coverage | Power | MSE |
| **200** | **MZIP** | -0.46 | | | 0.19 | -2.2 | | | 0.07 |
| | Delta | | 94.9% | 62.1% | 0.19 | | 95.0% | 61.8% | 0.07 |
| | Bootstrap | | 94.4% | 63.7% | 0.19 | | 94.5% | 64.1% | 0.07 |
| | **Poisson** | 24.26 | | | 0.51 | -12.25 | | | 0.1 |
| | Delta | | 21.9% | 90.0% | 0.07 | | 77.1% | 62.8% | 0.06 |
| | Bootstrap | | 89.6% | 40.0% | 0.35 | | 93.5% | 50.2% | 0.09 |
| **600** | **MZIP** | -0.61 | | | 0.11 | -0.27 | | | 0.04 |
| | Delta | | 95.2% | 98.6% | 0.11 | | 94.6% | 98.0% | 0.04 |
| | Bootstrap | | 94.6% | 98.6% | 0.11 | | 94.8% | 98.4% | 0.04 |
| | **Poisson** | 18.97 | | | 0.42 | -8.18 | | | 0.07 |
| | Delta | | 16.6% | 95.4% | 0.04 | | 73.2% | 97.2% | 0.04 |
| | Bootstrap | | 89.8% | 53.0% | 0.25 | | 93.0% | 88.8% | 0.05 |
| **1000** | **MZIP** | -0.05 | | | 0.08 | -0.67 | | | 0.03 |
| | Delta | | 94.8% | 100.0% | 0.08 | | 96.0% | 100.0% | 0.03 |
| | Bootstrap | | 94.8% | 100.0% | 0.08 | | 94.2% | 100.0% | 0.03 |
| | **Poisson** | 16.18 | | | 0.36 | -4.48 | | | 0.06 |
| | Delta | | 13.2% | 95.8% | 0.03 | | 70.6% | 99.6% | 0.03 |
| | Bootstrap | | 88.6% | 57.5% | 0.22 | | 92.8% | 94.2% | 0.04 |

**Web Table 4: Scenario 5 simulation results on a ratio scale**

| Sample Size | Outcome Model | NDE | | | | NIE | | | |
|---|---|---|---|---|---|---|---|---|---|
| | | Bias | Coverage | Power | MSE | Bias | Coverage | Power | MSE |
| **200** | **MZIP** | 1.68 | | | 0.21 | -3.0 | | | 0.07 |
| | Delta | | 95.0% | 52.8% | 0.21 | | 94.5% | 56.2% | 0.07 |
| | Bootstrap | | 94.4% | 54.3% | 0.21 | | 94.7% | 64.4% | 0.07 |
| | **Poisson** | 41.13 | | | 0.76 | -33.99 | | | 0.15 |
| | Delta | | 17.8% | 90.6% | 0.07 | | 57.8% | 54.0% | 0.05 |
| | Bootstrap | | 86.3% | 36.3% | 0.45 | | 84.7% | 21.0% | 0.09 |
| **600** | **MZIP** | -0.48 | | | 0.12 | -1.74 | | | 0.04 |
| | Delta | | 95.6% | 94.6% | 0.12 | | 96.0% | 98.3% | 0.04 |
| | Bootstrap | | 94.6% | 95.4% | 0.12 | | 93.8% | 97.9% | 0.04 |
| | **Poisson** | 33.49 | | | 0.58 | -24.1 | | | 0.10 |
| | Delta | | 14.0% | 96.4% | 0.04 | | 54.0% | 95.8% | 0.03 |
| | Bootstrap | | 86.0% | 48.6% | 0.33 | | 84.4% | 67.8% | 0.06 |
| **1000** | **MZIP** | -0.23 | | | 0.09 | -0.45 | | | 0.03 |
| | Delta | | 95.3% | 99.8% | 0.09 | | 95.2% | 99.8% | 0.03 |
| | Bootstrap | | 94.9% | 99.8% | 0.09 | | 94.6% | 99.9% | 0.03 |
| | **Poisson** | 29.33 | | | 0.57 | -20.97 | | | 0.08 |
| | Delta | | 13.2% | 96.2% | 0.03 | | 49.4% | 98.8% | 0.02 |
| | Bootstrap | | 86.6% | 53.2% | 0.28 | | 83.6% | 81.6% | 0.05 |

*Difference Scale*

**Web Table 5: Scenario 1 simulation results on a difference scale***

| Sample Size | Outcome Model | NDE | | | | NIE | | | |
|---|---|---|---|---|---|---|---|---|---|
| | | Bias | Coverage | Power | MSE | Bias | Coverage | Power | MSE |
| **200** | **MZIP** | 8.55 | | | 0.3 | 3.96 | | | 0.14 |
| | Delta | | 95.4% | 52.1% | 0.29 | | 94.6% | 52.3% | 0.14 |
| | Bootstrap | | 93.9% | 49.8% | 0.3 | | 95.1% | 65.4% | 0.14 |
| | **Poisson** | 32.91 | | | 0.66 | -4.03 | | | 0.21 |
| | Delta | | 39.4% | 82.9% | 0.17 | | 79.4% | 57.3% | 0.12 |
| | Bootstrap | | 90.6% | 38.5% | 0.46 | | 93.8% | 47.2% | 0.17 |
| | **ZIP** | -4.37 | | | 0.32 | 0.48 | | | 0.14 |
| | Quasi-Bayesian | | 95.3% | 32.7% | 0.33 | | 94.8% | 62.5% | 0.14 |
| | Bootstrap | | 94.5% | 45.1% | 0.32 | | 94.6% | 64.0% | 0.14 |
| **600** | **MZIP** | 9.05 | | | 0.18 | 4.32 | | | 0.08 |
| | Delta | | 93.8% | 92.2% | 0.17 | | 96.6% | 97.4% | 0.08 |
| | Bootstrap | | 92.8% | 92.0% | 0.17 | | 95.0% | 98.8% | 0.08 |
| | **Poisson** | 31.12 | | | 0.49 | 4.71 | | | 0.14 |
| | Delta | | 30.2% | 91.6% | 0.09 | | 76.4% | 97.0% | 0.07 |
| | Bootstrap | | 86.6% | 50.7% | 0.34 | | 94.6% | 90.6% | 0.11 |
| | **ZIP** | -7.21 | | | 0.2 | 3.67 | | | 0.08 |
| | Quasi-Bayesian | | 92.2% | 74.4% | 0.19 | | 95.0% | 98.0% | 0.08 |
| | Bootstrap | | 94.0% | 79.2% | 0.18 | | 94.6% | 98.2% | 0.08 |
| **1000** | **MZIP** | 8.08 | | | 0.13 | 5.21 | | | 0.06 |
| | Delta | | 93.3% | 99.2% | 0.13 | | 95.2% | 100.0% | 0.06 |
| | Bootstrap | | 93.4% | 99.0% | 0.13 | | 94.4% | 100.0% | 0.06 |
| | **Poisson** | 32.59 | | | 0.5 | 5.85 | | | 0.12 |
| | Delta | | 22.9% | 94.0% | 0.07 | | 71.8% | 99.7% | 0.06 |
| | Bootstrap | | 84.6% | 56.2% | 0.3 | | 95.2% | 94.0% | 0.09 |
| | **ZIP** | -9.16 | | | 0.15 | 3.19 | | | 0.06 |

|  | Quasi-Bayesian | 89.8% | 89.4% | 0.14 | 94.3% | 100.0% | 0.06 |
|  | Bootstrap | 95.0% | 92.5% | 0.14 | 94.6% | 100.0% | 0.06 |

*ZIP method uses simulation-based approach as opposed to the counterfactual approach to mediation and uses quasi-Bayesian or bootstrapping methods for standard error

**Effects are conditional on covariates, $c = 2$ was used because it is the mean of $\chi^2_2$

**Web Table 6: Scenario 2 simulation results on a difference scale***

| Sample Size | Outcome Model | NDE | | | | NIE | | | |
|---|---|---|---|---|---|---|---|---|---|
|  |  | Bias | Coverage | Power | MSE | Bias | Coverage | Power | MSE |
| **200** | **MZIP** | 7.4 |  |  | 0.35 | -1.59 |  |  | 0.16 |
|  | Delta |  | 93.9% | 43.1% | 0.34 |  | 92.4% | 29.2% | 0.16 |
|  | Bootstrap |  | 94.5% | 39.4% | 0.34 |  | 95.0% | 58.8% | 0.17 |
|  | **Poisson** | 44.23 |  |  | 1.04 | -34.71 |  |  | 0.29 |
|  | Delta |  | 30.7% | 83.9% | 0.17 |  | 59.2% | 48.5% | 0.1 |
|  | Bootstrap |  | 88.5% | 35.6% | 0.58 |  | 83.7% | 19.2% | 0.18 |
|  | **ZIP** | 15.87 |  |  | 0.4 | 16.41 |  |  | 0.19 |
|  | Quasi-Bayesian |  | 94.2% | 32.5% | 0.4 |  | 95.2% | 63.1% | 0.19 |
|  | Bootstrap |  | 92.6% | 44.0% | 0.38 |  | 94.2% | 58.8% | 0.19 |
| **600** | **MZIP** | 9.68 |  |  | 0.2 | 3.99 |  |  | 0.10 |
|  | Delta |  | 94.1% | 86.2% | 0.19 |  | 94.8% | 97.2% | 0.09 |
|  | Bootstrap |  | 93.6% | 83.0% | 0.19 |  | 93.5% | 98.6% | 0.09 |
|  | **Poisson** | 46.32 |  |  | 0.77 | -17.17 |  |  | 0.24 |
|  | Delta |  | 21.4% | 92.0% | 0.10 |  | 55.6% | 94.6% | 0.06 |
|  | Bootstrap |  | 86.2% | 44.6% | 0.45 |  | 87.0% | 68.6% | 0.13 |
|  | **ZIP** | 10.22 |  |  | 0.24 | 24.39 |  |  | 0.11 |
|  | Quasi-Bayesian |  | 92.8% | 70.8% | 0.23 |  | 90.4% | 97.6% | 0.11 |
|  | Bootstrap |  | 92.2% | 76.2% | 0.22 |  | 92.0% | 97.8% | 0.11 |
| **1000** | **MZIP** | 8.68 |  |  | 0.15 | 6.6 |  |  | 0.07 |
|  | Delta |  | 94.5% | 96.8% | 0.15 |  | 95.4% | 100.0% | 0.07 |
|  | Bootstrap |  | 94.3% | 96.8% | 0.15 |  | 94.0% | 100.0% | 0.07 |

| | | | | | | | | |
|---|---|---|---|---|---|---|---|---|
| | **Poisson** | 44.86 | | | 0.7 | -11.11 | | | 0.18 |
| | Delta | | 17.0% | 93.6% | 0.07 | | 51.6% | 97.8% | 0.05 |
| | Bootstrap | | 86.6% | 49.6% | 0.39 | | 88.6% | 81.4% | 0.11 |
| | **ZIP** | 6.42 | | | 0.18 | 26.37 | | | 0.09 |
| | Quasi-Bayesian | | 93.2% | 88.6% | 0.18 | | 87.0% | 100.0% | 0.09 |
| | Bootstrap | | 92.1% | 91.6% | 0.18 | | 89.0% | 100.0% | 0.08 |

*ZIP method uses simulation-based approach as opposed to the counterfactual approach to mediation and uses quasi-Bayesian or bootstrapping methods for standard error

**Effects are conditional on covariates, $c = 2$ was used because it is the mean of $\chi_2^2$

**Web Table 7: Scenario 3 simulation results on difference scale**

| Sample Size | Outcome Model | NDE | | | | NIE | | | |
|---|---|---|---|---|---|---|---|---|---|
| | | Bias | Coverage | Power | MSE | Bias | Coverage | Power | MSE |
| **200** | **MZIP** | 8.0 | | | 0.35 | 0.69 | | | 0.15 |
| | Delta | | 95.3% | 41.8% | 0.34 | | 93.7% | 37.5% | 0.15 |
| | Bootstrap | | 94.6% | 40.2% | 0.34 | | 94.8% | 60.5% | 0.15 |
| | **Poisson** | 17.81 | | | 0.79 | -10 | | | 0.25 |
| | Delta | | 37.0% | 79.2% | 0.17 | | 71.6% | 55.4% | 0.12 |
| | Bootstrap | | 92.8% | 27.2% | 0.55 | | 91.0% | 35.7% | 0.18 |
| | **ZIP** | 8.0 | | | 0.35 | 0.9 | | | 0.16 |
| | Quasi- Bayesian | | 95.1% | 35.1% | 0.37 | | 95.3% | 61.2% | 0.16 |
| | Bootstrap | | 93.6% | 41.2% | 0.35 | | 94.5% | 59.7% | 0.16 |
| **600** | **MZIP** | 8.45 | | | 0.2 | 4.34 | | | 0.09 |
| | Delta | | 94.3% | 83.3% | 0.2 | | 94.4% | 98.4% | 0.09 |
| | Bootstrap | | 94.6% | 86.5% | 0.2 | | 94.2% | 99.0% | 0.09 |
| | **Poisson** | 21.32 | | | 0.57 | -0.59 | | | 0.17 |
| | Delta | | 27.4% | 88.8% | 0.10 | | 68.8% | 96.8% | 0.07 |
| | Bootstrap | | 92.6% | 40.0% | 0.39 | | 94.0% | 84.4% | 0.12 |
| | **ZIP** | 5.15 | | | 0.2 | 5.55 | | | 0.09 |
| | Quasi-Bayesian | | 94.6% | 78.9% | 0.21 | | 95.6% | 97.4% | 0.09 |

|  |  | Bootstrap | 93.7% | 85.0% | 0.2 |  | 94.8% | 98.6% | 0.09 |
| --- | --- | --- | --- | --- | --- | --- | --- | --- | --- |
| 1000 | **MZIP** |  | 8.49 |  |  | 0.15 | 6.21 |  |  | 0.07 |
|  |  | Delta |  | 94.3% | 98.0% | 0.15 |  | 94.5% | 100.0% | 0.07 |
|  |  | Bootstrap |  | 93.4% | 97.4% | 0.15 |  | 95.6% | 100.0% | 0.07 |
|  | **Poisson** |  | 22.07 |  |  | 0.53 | 4.99 |  |  | 0.14 |
|  |  | Delta |  | 24.0% | 93.6% | 0.07 |  | 65.6% | 99.4% | 0.06 |
|  |  | Bootstrap |  | 91.4% | 49.0% | 0.35 |  | 96.2% | 90.2% | 0.1 |
|  | **ZIP** |  | 6.42 |  |  | 0.15 | 5.83 |  |  | 0.07 |
|  |  | Quasi-Bayesian |  | 94.2% | 96.0% | 0.15 |  | 93.3% | 100.0% | 0.07 |
|  |  | Bootstrap |  | 95.0% | 97.2% | 0.15 |  | 94.5% | 100.0% | 0.07 |

*ZIP method uses simulation-based approach as opposed to the counterfactual approach to mediation and uses quasi-Bayesian or bootstrapping methods for standard error

**Web Table 8: Scenario 4 simulation results on a difference scale**

| Sample Size | Outcome Model |  | NDE | | | | NIE | | | |
| --- | --- | --- | --- | --- | --- | --- | --- | --- | --- | --- |
|  |  |  | Bias | Coverage | Power | MSE | Bias | Coverage | Power | MSE |
| 200 | **MZIP** |  | 7.19 |  |  | 0.67 | 3 |  |  | 0.36 |
|  |  | Delta |  | 93.8% | 65.5% | 0.67 |  | 94.4% | 57.1% | 0.36 |
|  |  | Bootstrap |  | 95.0% | 62.6% | 0.68 |  | 95.2% | 65.2% | 0.37 |
|  | **Poisson** |  | 30.33 |  |  | 1.84 | -4.53 |  |  | 0.54 |
|  |  | Delta |  | 23.5% | 89.2% | 0.28 |  | 76.5% | 60.9% | 0.33 |
|  |  | Bootstrap |  | 89.1% | 39.1% | 1.20 |  | 94.0% | 49.1% | 0.45 |
|  | **ZIP** |  | -6.75 |  |  | 0.79 | -6.75 |  |  | 0.36 |
|  |  | Quasi- Bayesian |  | 92.0% | 45.1% | 0.72 |  | 95.2% | 66.8% | 0.36 |
|  |  | Bootstrap |  | 95.0% | 51.4% | 0.75 |  | 94.6% | 65.2% | 0.36 |
| 600 | **MZIP** |  | 8.32 |  |  | 0.4 | 5.55 |  |  | 0.2 |
|  |  | Delta |  | 93.7% | 98.2% | 0.38 |  | 95.0% | 97.7% | 0.21 |
|  |  | Bootstrap |  | 93.5% | 98.6% | 0.38 |  | 94.6% | 98.2% | 0.21 |

| | | | | | | | | |
|---|---|---|---|---|---|---|---|---|
| | **Poisson** | 31.61 | | | 1.42 | 2.26 | | 0.37 |
| | Delta | | 18.6% | 95.4% | 0.16 | | 75.2% | 97.3% | 0.19 |
| | Bootstrap | | 86.5% | 53.0% | 0.9 | | 93.3% | 88.8% | 0.29 |
| | **ZIP** | -12.02 | | | 0.49 | 3.68 | | | 0.2 |
| | Quasi-Bayesian | | 86.4% | 81.7% | 0.42 | | 94.6% | 98.0% | 0.2 |
| | Bootstrap | | 93.5% | 83.4% | 0.45 | | 94.7% | 98.2% | 0.2 |
| **1000** | **MZIP** | 8.25 | | | 0.3 | 6.05 | | | 0.16 |
| | Delta | | 92.8% | 100% | 0.29 | | 95.0% | 100.0% | 0.16 |
| | Bootstrap | | 90.7% | 100% | 0.3 | | 93.8% | 100.0% | 0.16 |
| | **Poisson** | 30.88 | | | 1.29 | 5.26 | | | 0.34 |
| | Delta | | 14.2% | 95.8% | 0.12 | | 69.2% | 99.7% | 0.15 |
| | Bootstrap | | 85.5% | 57.4% | 0.79 | | 94.5% | 94.1% | 0.23 |
| | **ZIP** | -15.27 | | | 0.4 | 2.72 | | | 0.16 |
| | Quasi-Bayesian | | 80.0% | 92.2% | 0.32 | | 94.2% | 100.0% | 0.15 |
| | Bootstrap | | 91.6% | 92.0% | 0.35 | | 95.1% | 99.9% | 0.16 |

*ZIP method uses simulation-based approach as opposed to the counterfactual approach to mediation and uses quasi-Bayesian or bootstrapping methods for standard error

**Web Table 9: Scenario 5 simulation results on a difference scale**

| Sample Size | Outcome Model | NDE | | | | NIE | | | |
|---|---|---|---|---|---|---|---|---|---|
| | | Bias | Coverage | Power | MSE | Bias | Coverage | Power | MSE |
| **200** | **MZIP** | 8.61 | | | 0.74 | 0.07 | | | 0.42 |
| | Delta | | 94.7% | 59.3% | 0.74 | | 94.1% | 41.9% | 0.41 |
| | Bootstrap | | 94.8% | 54.7% | 0.75 | | 94.5% | 64.6% | 0.42 |
| | **Poisson** | 37.52 | | | 2.74 | -34.58 | | | 0.87 |
| | Delta | | 20.2% | 90.2% | 0.28 | | 55.5% | 54.0% | 0.25 |
| | Bootstrap | | 89.1% | 36.3% | 1.51 | | 82.6% | 19.9% | 0.48 |
| | **ZIP** | 13.44 | | | 0.97 | 13.81 | | | 0.48 |

|  |  |  |  |  |  |  |  |  |
|---|---|---|---|---|---|---|---|---|
|  | Quasi- Bayesian |  | 92.7% | 43.9% | 0.91 |  | 94.4% | 62.9% | 0.5 |
|  | Bootstrap |  | 92.1% | 49.2% | 0.92 |  | 94.4% | 61.3% | 0.47 |
| **600** | **MZIP** | 7.9 |  |  | 0.43 | 5.58 |  |  | 0.24 |
|  | Delta |  | 93.9% | 96.1% | 0.42 |  | 96.4% | 98.3% | 0.24 |
|  | Bootstrap |  | 93.7% | 95.9% | 0.43 |  | 94.2% | 97.8% | 0.24 |
|  | **Poisson** | 43.11 |  |  | 2.11 | -16.7 |  |  | 0.56 |
|  | Delta |  | 13.7% | 96.3% | 0.16 |  | 54.2% | 95.8% | 0.16 |
|  | Bootstrap |  | 85.5% | 48.8% | 1.16 |  | 86.9% | 67.8% | 0.34 |
|  | **ZIP** | 5.2 |  |  | 0.61 | 22.4 |  |  | 0.29 |
|  | Quasi-Bayesian |  | 91.5% | 78.7% | 0.52 |  | 91.4% | 98.5% | 0.29 |
|  | Bootstrap |  | 92.8% | 82.6% | 0.55 |  | 91.7% | 98.5% | 0.28 |
| **1000** | **MZIP** | 8.06 |  |  | 0.33 | 6.26 |  |  | 0.18 |
|  | Delta |  | 93.6% | 99.9% | 0.33 |  | 94.6% | 99.9% | 0.18 |
|  | Bootstrap |  | 92.7% | 99.8% | 0.33 |  | 94.2% | 100.0% | 0.18 |
|  | **Poisson** | 43.05 |  |  | 1.92 | -11.83 |  |  | 0.5 |
|  | Delta |  | 13.5% | 96.5% | 0.13 |  | 49.3% | 98.9% | 0.12 |
|  | Bootstrap |  | 83.7% | 52.6% | 1.05 |  | 88.5% | 82.4% | 0.28 |
|  | **ZIP** | 1.49 |  |  | 0.5 | 24.56 |  |  | 0.22 |
|  | Quasi-Bayesian |  | 90.2% | 91.1% | 0.4 |  | 85.4% | 100.0% | 0.22 |
|  | Bootstrap |  | 94.0% | 89.6% | 0.43 |  | 90.6% | 100.0% | 0.22 |

*ZIP method uses simulation-based approach as opposed to the counterfactual approach to mediation and uses quasi-Bayesian or bootstrapping methods for standard error

The biggest takeaway from these simulations if that increasing the mean decreases coverage for Poisson and Scenario 3 really highlights the significance of zero-inflation attributable to the exposure and mediator on mediation effect estimation for Poisson model. Where appropriate fixed level of covariate was set to 2, the mean of the covariate's distribution.

In some scenarios, particularly those with higher differential effects in zero-inflated parameters, the ZIP method showed rather extreme and unpredictable bias. ZIP and MZIP make different model assumptions. ZIP assumes that the predictors have a linear relationship with the mean of the non-excess zeroes, whereas MZIP assumes the predictors have linear relationships with the overall mean (Liu et al., 2018). Our simulation was based on MZIP parameters such that MZIP should have little bias, but ZIP will be biased because of the assumption differences (Liu et al., 2018). We showed the simulation results here to show that this was considered, but ultimately, comparisons between ZIP and MZIP should not be made in terms of model parameters.

We did, however, see that MZIP exhibited more bias than expected in some scenarios on a difference scale. This is likely because while on a difference scale effects are conditional on covariates and since one of the covariates is a highly skewed distribution using average values for that fixed covariate may not approximate marginal mean estimates.

Web Appendix H: Simulations with Binary Mediators

Simulated data for binary mediators was performed in the same way as continuous mediators except the binary mediator is simulated based a random Bernoulli distribution, $M \sim Bernoulli\left(\frac{e^{\theta_0+\theta_1 x+\theta_4 c}}{1+e^{\theta_0+\theta_1 x+\theta_4 c}}\right)$. The following scenarios were considered based on the continuous mediator case.

**Web Table 10: Simulation scenarios considered for binary mediator model**

| Scenario | Outcome Model | Mediator Model | NDE | NIE |
|---|---|---|---|---|
| 1 | $\gamma=\{0.35,-1.5,0,0.25\}$ $\alpha=\{-0.6,0.41,0.6,0.25\}$ | Binary (Poisson) $\theta=\{0,2,0.25\}$ | IRR: 1.51 RD: 0.69 | IRR: 1.16 RD: 0.34 |
| 2 | $\gamma=\{0.35,-1.5,1.5,0.25\}$ $\alpha=\{-0.6,0.41,0.6,0.25\}$ | Binary (Poisson) $\theta=\{0,2,0.25\}$ | IRR: 1.51 RD: 0.69 | IRR: 1.16 RD: 0.34 |
| 3 | $\gamma=(0.35,-1.5,1.5,0.25)$ $\alpha=(0.4,0.41,0.6,0.25)$ | Binary (Poisson) $\theta=\{0,2,0.25\}$ | IRR: 1.51 RD: 1.88 | IRR: 1.16 RD: 0.92 |

Scenarios 1 and 2 are akin to Scenarios 1 and 2 for continuous mediator where Scenario 1 has high zero-inflation from exposure and Scenario 2 increases the amount of zero-inflation attributable from the mediator. Scenario 3 increases the overall mean from Scenario 2. Results are displayed in the following sections.

*Ratio Scale*

**Web Table 11: Scenario 1 (Binary) Simulation Results on a ratio scale**

| Sample Size | Outcome Model | NDE | | | | NIE | | | |
|---|---|---|---|---|---|---|---|---|---|
| | | Bias | Coverage | Power | MSE | Bias | Coverage | Power | MSE |
| **200** | **MZIP** | 0.92 | | | 0.23 | -1.78 | | | 0.07 |
| | Delta | | 94.8% | 45.6% | 0.22 | | 94.6% | 65.9% | 0.07 |
| | Bootstrap | | 94.0% | 50.2% | 0.23 | | 94.9% | 65.0% | 0.07 |
| | **Poisson** | 15.65 | | | 0.44 | -0.97 | | | 0.10 |
| | Delta | | 39.7% | 81.5% | 0.11 | | 74.2% | 79.3% | 0.05 |
| | Bootstrap | | 90.7% | 38.1% | 0.33 | | 94.4% | 45.8% | 0.08 |
| **600** | **MZIP** | -0.08 | | | 0.13 | -1.35 | | | 0.04 |
| | Delta | | 95.6% | 90.4% | 0.13 | | 94.9% | 98.3% | 0.04 |
| | Bootstrap | | 94.6% | 91.9% | 0.13 | | 95.1% | 98.2% | 0.04 |
| | **Poisson** | 9.54 | | | 0.32 | 1.22 | | | 0.06 |
| | Delta | | 33.4% | 91.8% | 0.06 | | 68.0% | 95.3% | 0.03 |
| | Bootstrap | | 91.6% | 54.1% | 0.22 | | 94.9% | 77.7% | 0.05 |
| **1000** | **MZIP** | 0.75 | | | 0.1 | -0.36 | | | 0.03 |
| | Delta | | 94.8% | 99.2% | 0.1 | | 94.5% | 100.0% | 0.03 |
| | Bootstrap | | 94.6% | 99.2% | 0.1 | | 94.7% | 100.0% | 0.03 |
| | **Poisson** | 10.62 | | | 0.27 | 0.42 | | | 0.05 |
| | Delta | | 30.0% | 94.8% | 0.05 | | 63.8% | 97.4% | 0.02 |
| | Bootstrap | | 92.4% | 66.0% | 0.19 | | 94.9% | 85.7% | 0.04 |

**Web Table 12: Scenario 2 (Binary) Simulation Results on a ratio scale**

| Sample Size | Outcome Model | NDE | | | | NIE | | | |
|---|---|---|---|---|---|---|---|---|---|
| | | Bias | Coverage | Power | MSE | Bias | Coverage | Power | MSE |
| **200** | **MZIP** | -0.71 | | | 0.36 | -2.47 | | | 0.08 |
| | Delta | | 95.1% | 22.1% | 0.34 | | 95.0% | 52.4% | 0.08 |
| | Bootstrap | | 95.5% | 19.3% | 0.37 | | 95.5% | 48.3% | 0.08 |
| | **Poisson** | 36.93 | | | 0.72 | -8.65 | | | 0.14 |
| | Delta | | 24.6% | 83.3% | 0.12 | | 71.1% | 71.4% | 0.05 |
| | Bootstrap | | 87.4% | 28.3% | 0.57 | | 92.5% | 19.9% | 0.11 |
| **600** | **MZIP** | 0.63 | | | 0.2 | -1.15 | | | 0.04 |
| | Delta | | 95.3% | 55.5% | 0.2 | | 95.2% | 92.0% | 0.04 |
| | Bootstrap | | 95.0% | 59.2% | 0.2 | | 94.4% | 88.2% | 0.04 |
| | **Poisson** | 20.57 | | | 0.52 | -1.55 | | | 0.09 |
| | Delta | | 22.9% | 90.0% | 0.06 | | 53.7% | 91.0% | 0.03 |
| | Bootstrap | | 89.9% | 40.0% | 0.37 | | 93.2% | 48.2% | 0.07 |
| **1000** | **MZIP** | 0.68 | | | 0.15 | -1.29 | | | 0.03 |
| | Delta | | 95.5% | 79.2% | 0.15 | | 95.1% | 98.8% | 0.03 |
| | Bootstrap | | 95.0% | 80.2% | 0.15 | | 94.5% | 98.5% | 0.03 |
| | **Poisson** | 18.38 | | | 0.44 | -1.75 | | | 0.07 |
| | Delta | | 18.6% | 93.2% | 0.05 | | 51.2% | 94.5% | 0.02 |
| | Bootstrap | | 89.9% | 44.4% | 0.3 | | 93.3% | 66.1% | 0.05 |

**Web Table 13: Scenario 3 (Binary) Simulation Results on a ratio scale**

| Sample Size | Outcome Model | NDE | | | | NIE | | | |
|---|---|---|---|---|---|---|---|---|---|
| | | Bias | Coverage | Power | MSE | Bias | Coverage | Power | MSE |
| **200** | **MZIP** | -4.23 | | | 0.3 | -2.37 | | | 0.07 |
| | Delta | | 94.6% | 27.9% | 0.29 | | 94.3% | 67.3% | 0.06 |
| | Bootstrap | | 95.6% | 26.3% | 0.31 | | 95.1% | 62.3% | 0.07 |
| | **Poisson** | 37.77 | | | 0.76 | -11.63 | | | 0.14 |
| | Delta | | 18.4% | 89.8% | 0.07 | | 53.6% | 82.7% | 0.04 |
| | Bootstrap | | 87.5% | 29.6% | 0.56 | | 91.1% | 20.6% | 0.11 |
| **600** | **MZIP** | -1.57 | | | 0.17 | -1.05 | | | 0.04 |
| | Delta | | 94.6% | 67.8% | 0.17 | | 94.7% | 98.1% | 0.04 |
| | Bootstrap | | 94.9% | 70.6% | 0.17 | | 95.4% | 96.6% | 0.04 |
| | **Poisson** | 21.4 | | | 0.51 | -3.04 | | | 0.02 |
| | Delta | | 13.8% | 93.4% | 0.04 | | 46.2% | 95.0% | 0.07 |
| | Bootstrap | | 89.4% | 37.7% | 0.35 | | 93.5% | 49.7% | 0.07 |
| **1000** | **MZIP** | -1.08 | | | 0.13 | -0.33 | | | 0.03 |
| | Delta | | 94.8% | 90.5% | 0.13 | | 94.9% | 100% | 0.03 |
| | Bootstrap | | 94.4% | 89.4% | 0.13 | | 93.6% | 99.8% | 0.03 |
| | **Poisson** | 15.47 | | | 0.47 | -1.25 | | | 0.07 |
| | Delta | | 11.2% | 96.1% | 0.03 | | 43.9% | 97.4% | 0.02 |
| | Bootstrap | | 89.0% | 42.5% | 0.3 | | 92.7% | 67.1% | 0.05 |

These results reinforce results from the case with a continuous mediator, that is Poisson is biased and has poor coverage especially when $\gamma_2$ is increased.

Difference Scale

**Web Table 14: Scenario 1 (Binary) Simulation Results on a difference scale**

| Sample Size | Outcome Model | NDE | | | | NIE | | | |
|---|---|---|---|---|---|---|---|---|---|
| | | Bias | Coverage | Power | MSE | Bias | Coverage | Power | MSE |
| **200** | **MZIP** | -0.46 | | | 0.37 | -3.68 | | | 0.14 |
| | Delta | | 97.7% | 43.3% | 0.39 | | 85.9% | 68.8% | 0.15 |
| | Bootstrap | | 94.8% | 49.5% | 0.36 | | 95.2% | 65.7% | 0.14 |
| | **Poisson** | 9.65 | | | 0.7 | 20.70 | | | 0.31 |
| | Delta | | 44.1% | 81.5% | 0.18 | | 72.6% | 77.2% | 0.15 |
| | Bootstrap | | 92.1% | 38.2% | 0.52 | | 93.5% | 47.4% | 0.28 |
| | **ZIP** | -6.06 | | | 0.36 | -0.59 | | | 0.13 |
| | Quasi- Bayesian | | 93.9% | 36.7% | 0.37 | | 97.7% | 83.0% | 0.14 |
| | Bootstrap | | 94.4% | 46.5% | 0.37 | | 95.9% | 87.4% | 0.14 |
| **600** | **MZIP** | -1.96 | | | 0.21 | -1.94 | | | 0.08 |
| | Delta | | 96.4% | 92.1% | 0.22 | | 90.7% | 100% | 0.09 |
| | Bootstrap | | 93.4% | 91.2% | 0.21 | | 94.5% | 98.4% | 0.08 |
| | **Poisson** | 5.82 | | | 0.51 | 21.99 | | | 0.18 |
| | Delta | | 35.4% | 91.3% | 0.1 | | 60.6% | 95.5% | 0.08 |
| | Bootstrap | | 93.1% | 52.2% | 0.36 | | 89.8% | 77.6% | 0.15 |
| | **ZIP** | -14.91 | | | 0.23 | 0.77 | | | 0.07 |
| | Quasi-Bayesian | | 89.4% | 75.2% | 0.21 | | 96.6% | 100% | 0.08 |
| | Bootstrap | | 92.7% | 80.9% | 0.21 | | 97.1% | 100% | 0.08 |
| **1000** | **MZIP** | 0.54 | | | 0.16 | -0.81 | | | 0.06 |
| | Delta | | 96.0% | 99.3% | 0.17 | | 94.1% | 100.0% | 0.07 |
| | Bootstrap | | 94.6% | 99.1% | 0.16 | | 95.0% | 100.0% | 0.06 |
| | **Poisson** | 9.36 | | | 0.42 | 23.11 | | | 0.15 |
| | Delta | | 29.2% | 94.8% | 0.08 | | 52.2% | 97.2% | 0.06 |
| | Bootstrap | | 92.9% | 64.2% | 0.3 | | 88.2% | 86.0% | 0.13 |
| | **ZIP** | -18.7 | | | 0.18 | 1.72 | | | 0.06 |

|  |  | 83.6% | 90.2% | 0.16 | 96.7% | 100.0% | 0.06 |
| Quasi-Bayesian |
|  | Bootstrap | 92.2% | 92.5% | 0.17 | 96.5% | 100.0% | 0.06 |

*ZIP method uses simulation-based approach as opposed to the counterfactual approach to mediation and uses quasi-Bayesian or bootstrapping methods for standard error

**Web Table 15: Scenario 2 (Binary) Simulation Results on a difference scale**

| Sample Size | Outcome Model | NDE | | | | NIE | | | |
|---|---|---|---|---|---|---|---|---|---|
|  |  | Bias | Coverage | Power | MSE | Bias | Coverage | Power | MSE |
| **200** | **MZIP** | -1.25 |  |  | 0.54 | -5.77 |  |  | 0.18 |
|  | Delta |  | 96.7% | 14.3% | 0.57 |  | 82.5% | 32.4% | 0.18 |
|  | Bootstrap |  | 95.9% | 20.2% | 0.58 |  | 95.0% | 48.5% | 0.19 |
|  | **Poisson** | 27.52 |  |  | 1.10 | 7.02 |  |  | 0.41 |
|  | Delta |  | 28.1% | 81.0% | 0.19 |  | 57.2% | 69.3% | 0.15 |
|  | Bootstrap |  | 89.6% | 30.0% | 0.83 |  | 93.5% | 20.4% | 0.36 |
|  | **ZIP** | -12.1 |  |  | 0.58 | 30.22 |  |  | 0.17 |
|  | Quasi- Bayesian |  | 95.0% | 17.0% | 0.58 |  | 93.1% | 88.9% | 0.18 |
|  | Bootstrap |  | 94.6% | 25.9% | 0.59 |  | 95.7% | 77.6% | 0.19 |
| **600** | **MZIP** | -0.71 |  |  | 0.32 | -2.13 |  |  | 0.11 |
|  | Delta |  | 96.8% | 55.7% | 0.32 |  | 89.6% | 100% | 0.1 |
|  | Bootstrap |  | 92.5% | 57.2% | 0.31 |  | 93.7% | 88.4% | 0.1 |
|  | **Poisson** | 12.85 |  |  | 0.80 | 22.97 |  |  | 0.24 |
|  | Delta |  | 23.4% | 89.4% | 0.11 |  | 49.4% | 90.3% | 0.08 |
|  | Bootstrap |  | 91.6% | 35.8% | 0.57 |  | 93.5% | 49.3% | 0.21 |
|  | **ZIP** | -11.56 |  |  | 0.33 | 34.98 |  |  | 0.1 |
|  | Quasi-Bayesian |  | 93.5% | 44.7% | 0.32 |  | 76.9% | 100% | 0.1 |
|  | Bootstrap |  | 94.6% | 50.9% | 0.32 |  | 81.0% | 99.8% | 0.1 |
| **1000** | **MZIP** | -0.63 |  |  | 0.24 | -1.42 |  |  | 0.08 |
|  | Delta |  | 95.8% | 82.7% | 0.25 |  | 90.8% | 100.0% | 0.08 |

|  |  | Bootstrap | 95.5% | 78.2% | 0.24 |  | 94.8% | 98.2% | 0.08 |
|---|---|---|---|---|---|---|---|---|---|
|  | **Poisson** |  | 17.52 |  |  | 0.69 | 25.32 |  |  | 0.2 |
|  |  | Delta |  | 20.7% | 93.7% | 0.08 |  | 45.1% | 95.1% | 0.06 |
|  |  | Bootstrap |  | 90.9% | 44.5% | 0.46 |  | 92.0% | 68.0% | 0.17 |
|  | **ZIP** |  | -11.62 |  |  | 0.25 | 36.80 |  |  | 0.08 |
|  |  | Quasi-Bayesian |  | 92.3% | 65.4% | 0.25 |  | 62.9% | 100.0% | 0.08 |
|  |  | Bootstrap |  | 94.4% | 70.2% | 0.25 |  | 61.8% | 100.0% | 0.08 |

*ZIP method uses simulation-based approach as opposed to the counterfactual approach to mediation and uses quasi-Bayesian or bootstrapping methods for standard error

**Web Table 16: Scenario 3 (Binary) Simulation Results on a difference scale**

| Sample Size | Outcome Model |  | NDE | | | | NIE | | | |
|---|---|---|---|---|---|---|---|---|---|---|
|  |  |  | Bias | Coverage | Power | MSE | Bias | Coverage | Power | MSE |
| **200** | **MZIP** |  | -3.77 |  |  | 1.27 | -5.68 |  |  | 0.42 |
|  |  | Delta |  | 95.5% | 25.5% | 1.29 |  | 83.7% | 80.3% | 0.38 |
|  |  | Bootstrap |  | 95.2% | 25.4% | 1.34 |  | 94.8% | 61.3% | 0.44 |
|  | **Poisson** |  | 27.18 |  |  | 2.88 | 3.57 |  |  | 1.03 |
|  |  | Delta |  | 16.8% | 89.1% | 0.32 |  | 47.5% | 82.3% | 0.3 |
|  |  | Bootstrap |  | 90.3% | 29.2% | 2.20 |  | 93.2% | 19.2% | 0.95 |
|  | **ZIP** |  | -14.59 |  |  | 1.43 | 30.25 |  |  | 0.44 |
|  |  | Quasi- Bayesian |  | 93.3% | 18.1% | 1.44 |  | 92.1% | 95.6% | 0.45 |
|  |  | Bootstrap |  | 93.9% | 27.9% | 1.47 |  | 95.1% | 84.7% | 0.49 |
| **600** | **MZIP** |  | -2.32 |  |  | 0.72 | -0.79 |  |  | 0.24 |
|  |  | Delta |  | 96.2% | 72.1% | 0.74 |  | 88.6% | 100% | 0.23 |
|  |  | Bootstrap |  | 94.7% | 70.5% | 0.71 |  | 95.1% | 98.0% | 0.24 |
|  | **Poisson** |  | 17.67 |  |  | 2.14 | 19.17 |  |  | 0.66 |
|  |  | Delta |  | 13.6% | 95.0% | 0.18 |  | 37.4% | 93.9% | 0.16 |
|  |  | Bootstrap |  | 90.4% | 36.1% | 1.51 |  | 94.9% | 50.1% | 0.56 |

|  |  |  |  |  |  |  |  |  |  |
|---|---|---|---|---|---|---|---|---|---|
|  | ZIP | -14.76 |  |  | 0.83 | 34.93 |  |  | 0.25 |
|  | Quasi-Bayesian |  | 93.0% | 46.0% | 0.81 |  | 71.8% | 100% | 0.26 |
|  | Bootstrap |  | 94.9% | 52.6% | 0.83 |  | 79.3% | 100% | 0.26 |
| **1000** | **MZIP** | -0.35 |  |  | 0.55 | -0.79 |  |  | 0.18 |
|  | Delta |  | 95.2% | 92.7% | 0.56 |  | 90.8% | 100.0% | 0.18 |
|  | Bootstrap |  | 93.3% | 88.2% | 0.55 |  | 94.1% | 99.9% | 0.19 |
|  | **Poisson** | 21.46 |  |  | 1.88 | 21.46 |  |  | 0.56 |
|  | Delta |  | 12.2% | 96.2% | 0.14 |  | 36.2% | 96.7% | 0.13 |
|  | Bootstrap |  | 89.2% | 43.7% | 1.31 |  | 91.4% | 67.8% | 0.44 |
|  | **ZIP** | -18.45 |  |  | 0.66 | 36.78 |  |  | 0.19 |
|  | Quasi-Bayesian |  | 89.9% | 65.5% | 0.62 |  | 54.7% | 100.0% | 0.20 |
|  | Bootstrap |  | 93.7% | 68.2% | 0.65 |  | 61.4% | 100.0% | 0.20 |

For binary mediators on a difference scale, MZIP is mostly unbiased, but has coverage that is slightly inadequate for NIE when using the delta method. NIE when using ZIP became more biased when the zero-inflation caused by the mediator increased and when the overall mean increased and also had poor coverage for both bootstrap and quasi-Bayesian error approaches. ZIP did somewhat better for NDE, but still was not optimal.

Web Appendix I: Simulations with over-dispersed zero-inflated count mediators

Because of the limitations of Poisson models, we also considered simulations for an over-dispersed zero-inflated count outcome. In this section we consider the same nominal parameter values from Scenario 2 in the previous sections, but instead of simulating a zero-inflated Poisson distribution for the outcome, we simulate a zero-inflated negative binomial outcome.

$$\psi \sim Bernoulli\left(\frac{\exp(Z\gamma)}{1 + \exp(Z\gamma)}\right)$$

$$\mu \sim Negative\ Binomial(\exp(Z\alpha + \log(1 + Z\gamma)), \omega = 2)$$

Where outcome $Y = (1 - \psi)\mu$ and $\omega$ is a dispersion parameter. The mean $E(\mu) = \exp(Z\alpha + \log(1 + Z\gamma))$, but the variance is $Var(\mu) = E(\mu) + \left(\frac{E(\mu)^2}{\omega}\right)$. Results on a ratio scale are shown in Table A9.1 and on a difference scale in Table A9.2.

For the case with over-dispersion, we notice that on the ratio scale bias increase slightly for MZIP but decreases slightly for Poisson likely because the over-dispersion lessens the impact of zero-inflation. Use of delta method with model-based covariance matrix results in slight underestimation of variance and poor coverage for MZIP. Use of a robust covariance matrix results in MZIP having coverage near the nominal level, but Poisson is still slightly below nominal. Bootsrap standard errors had near nominal coverage for MZIP but can be computationally intensive so it may be advisable to use robust delta method instead. Even with robust standard errors or bootstrapping Poisson will give more biased results than MZIP and thus should be avoided with zero-inflated outcomes.

On a difference scale, the Poisson outcome model gave biased estimates of NIE and NDE similar to the Scenario 2. While the trends on the ratio and difference scale are similar, on a difference scale coverage was lower for both MZIP and Poisson when using robust variance estimates and bootstrapping than on the ratio scale. MZIP was still near the nominal 95% coverage using these variance estimation methods. Using the simulation-based approach utilizing ZIP our estimates of NDE were less biased than when using MZIP, but our NIE estimates were more biased. More investigation is needed to determine why this occurred. With the over-dispersion we also witnessed an under-estimation of variance, although bootstrapping coverage was only slightly below nominal.

While the results using the simulation-based approach were inconsistent, we can argue that the proposed method using MZIP performs well with over-dispersed zero-inflated counts when using bootstrapped standard errors or robust covariance delta method. It also seems that overall, results on a ratio scale are more reliable and consistent than on a difference scale, potentially due to the difference scale requiring more conditionalities. The simulation-based approach also allows for zero-inflated negative binomial (ZINB) outcome models to be used. We omit comparison using ZINB until the proposed methodology has been extended to marginalized zero-inflated negative binomial model

**Web Table 17: Simulation results for over-dispersed zero-inflated outcome on a ratio scale***

| Sample Size | Outcome Model | NDE | | | | NIE | | | |
|---|---|---|---|---|---|---|---|---|---|
| | | % Bias | Coverage | Power | MSE | % Bias | Coverage | Power | MSE |
| **200** | **MZIP** | 8.3 | | | 0.30 | -10.13 | | | 0.08 |
| | Delta | | 89.5% | 46.1% | 0.24 | | 89.6% | 49.1% | 0.07 |
| | Robust | | 94.1% | 36.1% | 0.28 | | 91.1% | 40.3% | 0.08 |
| | Bootstrap | | 94.7% | 36.6% | 0.3 | | 94.3% | 56.3% | 0.08 |
| | **Poisson** | 32.3 | | | 0.6 | -16.67 | | | 0.12 |
| | Delta | | 32.0% | 83.5% | 0.11 | | 73.0% | 55.5% | 0.06 |
| | Robust | | 92.0% | 28.8% | 0.42 | | 89.6% | 16.8% | 0.09 |
| | Bootstrap | | 90.0% | 33.4% | 0.41 | | 92.3% | 43.1% | 0.09 |
| **600** | **MZIP** | 2.21 | | | 0.17 | -3.34 | | | 0.05 |
| | Delta | | 89.6% | 87.4% | 0.14 | | 91.0% | 98.0% | 0.04 |
| | Robust | | 94.7% | 77.3% | 0.16 | | 94.0% | 96.4% | 0.04 |
| | Bootstrap | | 94.4% | 77.0% | 0.17 | | 94.2% | 97.5% | 0.05 |
| | **Poisson** | 19.86 | | | 0.46 | -5.04 | | | 0.08 |
| | Delta | | 25.8% | 92.5% | 0.06 | | 71.8% | 96.3% | 0.04 |
| | Robust | | 91.8% | 42.0% | 0.29 | | 91.8% | 72.2% | 0.06 |
| | Bootstrap | | 90.0% | 46.4% | 0.29 | | 92.2% | 86.8% | 0.06 |
| **1000** | **MZIP** | 2.06 | | | 0.14 | -3.97 | | | 0.04 |
| | Delta | | 86.6% | 94.6% | 0.11 | | 92.6% | 100.0% | 0.03 |
| | Robust | | 95.3% | 91.3% | 0.13 | | 94.2% | 100.0% | 0.03 |
| | Bootstrap | | 94.1% | 90.2% | 0.13 | | 94.5% | 99.7% | 0.04 |
| | **Poisson** | 20.25 | | | 0.46 | -6.05 | | | 0.07 |
| | Delta | | 19.4% | 93.0% | 0.05 | | 68.0% | 99.6% | 0.03 |
| | Robust | | 90.0% | 48.4% | 0.25 | | 92.4% | 82.6% | 0.05 |
| | Bootstrap | | 89.6% | 55.0% | 0.25 | | 92.2% | 91.4% | 0.05 |

*Bias is percent median bias computed as (true value-median estimated)/true value, MSE is the median standard error with the intrinsic value being computed as the standard error between simulated estimates.

**Web Table 18: Simulation results for over-dispersed zero-inflated outcome on a difference scale***

| Sample Size | Outcome Model | NDE | | | | NIE | | | |
|---|---|---|---|---|---|---|---|---|---|
| | | Bias | Coverage | Power | MSE | Bias | Coverage | Power | MSE |
| **200** | **MZIP** | 10.79 | | | 0.43 | -10.14 | | | 0.19 |
| | Delta | | 88.9% | 45.8% | 0.35 | | 86.5% | 23.0% | 0.15 |
| | Robust | | 93.4% | 40.3% | 0.41 | | 92.1% | 33.1% | 0.17 |
| | Bootstrap | | 94.6% | 33.7% | 0.42 | | 93.1% | 44.7% | 0.18 |
| | **Poisson** | 41.19 | | | 1.19 | -38.45 | | | 0.47 |
| | Delta | | 27.9% | 84.3% | 0.17 | | 53.7% | 46.1% | 0.09 |
| | Robust | | 92.2% | 29.3% | 0.66 | | 75.6% | 0.9% | 0.17 |
| | Bootstrap | | 89.7% | 32.0% | 0.64 | | 81.1% | 15.8% | 0.19 |
| | **ZIP** | 2.99 | | | 0.50 | 7.85 | | | 0.21 |
| | Quasi- Bayesian | | 88.0% | 31.3% | 0.41 | | 91.0% | 58.3% | 0.19 |
| | Bootstrap | | 94.1% | 32.2% | 0.46 | | 93.2% | 46.6% | 0.19 |
| **600** | **MZIP** | 13.11 | | | 0.25 | -0.89 | | | 0.12 |
| | Delta | | 87.8% | 79.8% | 0.20 | | 88.1% | 96.4% | 0.09 |
| | Robust | | 93.2% | 75.3% | 0.24 | | 94.6% | 96.0% | 0.11 |
| | Bootstrap | | 94.2% | 68.6% | 0.24 | | 94.4% | 97.6% | 0.10 |
| | **Poisson** | 42.29 | | | 1.28 | -19.56 | | | 0.25 |
| | Delta | | 20.6% | 93.0% | 0.10 | | 49.2% | 93.0% | 0.06 |
| | Robust | | 85.6% | 45.0% | 0.48 | | 81.1% | 24.8% | 0.13 |
| | Bootstrap | | 88.2% | 43.0% | 0.48 | | 87.4% | 60.1% | 0.13 |
| | **ZIP** | 1.61 | | | 0.34 | 17.57 | | | 0.13 |
| | Quasi- Bayesian | | 83.2% | 58.6% | 0.23 | | 86.4% | 98.0% | 0.11 |
| | Bootstrap | | 93.2% | 52.2% | 0.29 | | 95.2% | 97.2% | 0.12 |
| **1000** | **MZIP** | 10.01 | | | 0.19 | 2.41 | | | 0.09 |
| | Delta | | 87.8% | 93.4% | 0.16 | | 90.6% | 99.9% | 0.07 |
| | Robust | | 94.1% | 92.4% | 0.17 | | 95.7% | 99.6% | 0.08 |
| | Bootstrap | | 94.1% | 86.4% | 0.19 | | 95.4% | 99.6% | 0.08 |
| | **Poisson** | 47.62 | | | 0.73 | -11.92 | | | 0.22 |

| | | | | | | | |
|---|---|---|---|---|---|---|---|
| Delta | | 19.0% | 94.2% | 0.07 | 48.2% | 98.4% | 0.05 |
| Robust | | 86.2% | 48.6% | 0.44 | 84.8% | 52.0% | 0.11 |
| Bootstrap | | 84.6% | 48.6% | 0.42 | 86.6% | 76.6% | 0.11 |
| **ZIP** | 2.30 | | | 0.30 | 21.43 | | 0.10 |
| Quasi- Bayesian | | 77.0% | 74.6% | 0.18 | 84.0% | 99.9% | 0.08 |
| Bootstrap | | 92.6% | 62.6% | 0.23 | 92.2% | 100% | 0.09 |

*ZIP method uses simulation-based approach as opposed to the counterfactual approach to mediation and uses quasi-Bayesian or bootstrapping methods for standard error

**For formulas conditional on covariates, $c = 2$ was used because it is the mean of $\chi^2_2$

Web Appendix J: Application results on a difference scale

The application depicted in Section 6 of the manuscript was also completed on a difference scale and is shown in Tables S19. The ZIP method was used with the R package 'maczic' (Cheng et al., 2018; Cheng, Guo and Cheng, 2021; R Core Team, 2020). 1000 resamples are used for quasi-Bayesian and bootstrapped variance estimates

Using ZIP with quasi-Bayesian standard errors, all hypotheses were the same as MZIP with robust standard errors. The quasi-Bayesian standard errors were noticeably larger than the MZIP model with robust standard errors, but in the bootstrap section (Table S21) we see that both the ZIP and MZIP method have similar bootstrap variance estimates which are similar to the robust standard errors with MZIP. Looking into the model reveals that some of the parameter estimates in the ZIP model had larger variance estimates than expected although there was no evidence of separations of multicollinearity. It seems that this larger variance may be creating some instability in the approximated posterior distribution's estimates of NIE and NDE (King, Tomz and Wittenberg, 2000).

**Web Table 19: Results showing how baseline BMI differences in the number of inpatient care visits 2 years after baseline can be explained by baseline diabetes status using REGARDS (2003-2007).**

| Outcome Model | Estimate | 18.5-<25 kg/m$^2$ (n=2613) | 25-<30 kg/m$^2$ (n=3729) | 30-<35 kg/m$^2$ (n=2047) | 35-<40 kg/m$^2$ (n=775) | ≥40 kg/m$^2$ (n=484) |
|---|---|---|---|---|---|---|
| **MZIP*** (model) | NIE | Ref | 0.024(0.019,0.030) | 0.054(0.046,0.063) | 0.080(0.065,0.096) | 0.109(0.085,0.133) |
| | NDE | Ref | 0.008(-0.038,0.053) | 0.050(-0.004,0.103) | 0.036(-0.039,0.111) | 0.111(0.011,0.212) |
| | TE | Ref | 0.032(-0.014,0.077) | 0.104(0.048,0.159) | 0.116(0.033,0.199) | 0.219(0.103,0.335) |
| | PM | Ref | 76% | 53% | 69% | 50% |
| **MZIP*** (robust) | NIE | Ref | 0.024(0.019,0.030) | 0.054(0.046,0.063) | 0.080(0.064,0.096) | 0.109(0.083,0.136) |
| | NDE | Ref | 0.008(-0.043,0.058) | 0.050(-0.010,0.108) | 0.036(-0.048,0.119) | 0.111(-0.011,0.230) |
| | TE | Ref | 0.032(-0.020,0.083) | 0.104(0.042,0.165) | 0.116(0.022,0.209) | 0.219(0.079,0.360) |
| | PM | ref | 76% | 53% | 69% | 50% |
| **Poisson*** | NIE | Ref | 0.025(0.019,0.031) | 0.055(0.047,0.064) | 0.082(0.068,0.096) | 0.111(0.089,0.133) |
| | NDE | Ref | 0.006(-0.030,0.042) | 0.040(-0.003,0.082) | 0.030(-0.028,0.088) | 0.101(0.026,0.177) |
| | TE | Ref | 0.030(-0.006,0.067) | 0.094(0.050,0.139) | 0.103(0.047,0.176) | 0.211(0.124,0.298) |
| | PM | Ref | 82% | 59% | 74% | 53% |
| **ZIP**** | NIE | Ref | 0.025(0.017,0.035) | 0.055(0.038,0.072) | 0.083(0.057,0.113) | 0.112(0.077,0.151) |
| | NDE | Ref | 0.007(-0.061,0.080) | 0.048(-0.032,0.128) | 0.040(-0.067,0.149) | 0.114(-0.017,0.263) |
| | TE | Ref | 0.032(-0.037,0.105) | 0.103(0.019,0.183) | 0.122(0.007,0.245) | 0.226(0.080,0.390) |
| | PM | Ref | 78% | 53% | 68% | 49% |

*Standard errors computed using delta method

***Mean values for each covariate were used to estimate mediation effects. Delta method used for variance estimation for MZIP and Poisson. Quasi-Bayesian with 1000 simulations used for ZIP method.

## Web Appendix K: Application with Bootstrap Standard Errors

**Web Table 20: Results showing how BMI differences in the number of inpatient care visits can be explained by diabetes with bootstrapped standard errors (ratio scale).**

| Outcome Model | Estimate | 18.5-<25 kg/m$^2$ (n=2613) | 25-<30 kg/m$^2$ (n=3729) | 30-<35 kg/m$^2$ (n=2047) | 35-<40 kg/m$^2$ (n=775) | ≥40 kg/m$^2$ (n=484) |
|---|---|---|---|---|---|---|
| **MZIP** | IRRNIE | Ref | 1.05 (1.04,1.07) | 1.10 (1.08,1.13) | 1.16 (1.12,1.20) | 1.19 (1.14,1.24) |
|  | IRRNDE | Ref | 1.02 (0.91,1.12) | 1.10 (0.98,1.23) | 1.07 (0.91,1.26) | 1.23 (0.99,1.51) |
|  | IRRTE | Ref | 1.07 (0.95,1.18) | 1.22 (1.08,1.36) | 1.24 (1.06,1.46) | 1.46 (1.18,1.77) |
|  | PM | Ref | 76% | 53% | 69% | 50% |
| **Poisson** | IRRNIE | Ref | 1.05 (1.04,1.07) | 1.11 (1.08,1.13) | 1.16 (1.12,1.20) | 1.19 (1.15,1.25) |
|  | IRRNDE | Ref | 1.01 (0.91,1.13) | 1.08 (0.97,1.22) | 1.06 (0.90,1.24) | 1.21 (0.98,1.46) |
|  | IRRTE | Ref | 1.06 (0.96,1.18) | 1.20 (1.07,1.34) | 1.23 (1.05,1.44) | 1.44 (1.16,1.74) |
|  | PM | Ref | 82% | 59% | 74% | 53% |

*Standard error computed using delta method

**Web Table 21: Results showing how BMI differences in the number of inpatient care visits can be explained by diabetes with bootstrap standard errors (difference scale).**

| Outcome Model | Estimate | 18.5-<25 kg/m² (n=2613) | 25-<30 kg/m² (n=3729) | 30-<35 kg/m² (n=2047) | 35-<40 kg/m² (n=775) | ≥40 kg/m² (n=484) |
|---|---|---|---|---|---|---|
| **MZIP** | NIE | Ref | 0.024(0.018,0.031) | 0.054(0.041,0.069) | 0.080(0.059,0.100) | 0.109(0.079,0.145) |
| | NDE | Ref | 0.008(-0.043,0.056) | 0.050(-0.014,0.112) | 0.036(-0.050,0.122) | 0.111(-0.018,0.230) |
| | TE | Ref | 0.032(-0.022,0.083) | 0.104(0.040,0.166) | 0.116(0.022,0.209) | 0.219(0.091,0.353) |
| | PM | Ref | 76% | 53% | 69% | 50% |
| **Poisson** | NIE | Ref | 0.025(0.018,0.033) | 0.055(0.042,0.071) | 0.082(0.061,0.107) | 0.111(0.080,0.151) |
| | NDE | Ref | 0.006(-0.045,0.055) | 0.040(-0.018,0.094) | 0.030(-0.052,0.116) | 0.101(-0.007,0.229) |
| | TE | Ref | 0.030(-0.022,0.081) | 0.094(0.036,0.153) | 0.103(0.022,0.212) | 0.211(0.090,0.362) |
| | PM | Ref | 82% | 59% | 74% | 53% |
| **ZIP** | NIE | Ref | 0.025(0.018,0.034) | 0.055(0.041,0.071) | 0.083(0.061,0.107) | 0.112(0.078,0.152) |
| | NDE | Ref | 0.007(-0.045,0.060) | 0.048(-0.013,0.107) | 0.040(-0.051,0.140) | 0.114(0.001,0.235) |
| | TE | Ref | 0.032(-0.021,0.085) | 0.103(0.041,0.163) | 0.122(0.033,0.212) | 0.226(0.096,0.366) |
| | PM | Ref | 78% | 53% | 68% | 49% |

\* Mean values of each covariate were used to estimate mediation effects. Delta method used for variance estimation for MZIP and Poisson. Quasi-Bayesian with 1000 simulations used for ZIP method.

Web Appendix L: Forest Plots of Application Results

In this section we show forest plots of the results for the application on both a ratio and difference scale. Based on these figures it is easy to see that Poisson model has narrower confidence intervals than the proposed method. However, using robust standard errors for the MZIP model we obtain noticeably wider confidence intervals than when using model-based errors. So, it seems there is some over-dispersion occurring.

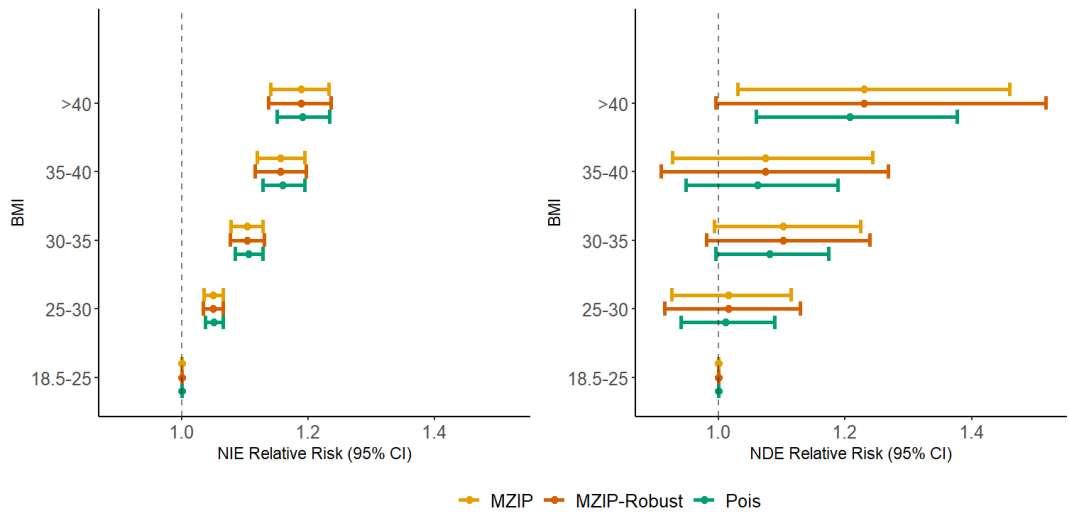

*Web Figure 1: Forest plot comparing IRRNIE and IRRNDE for Poisson vs. MZIP in the data application using REGARDS*

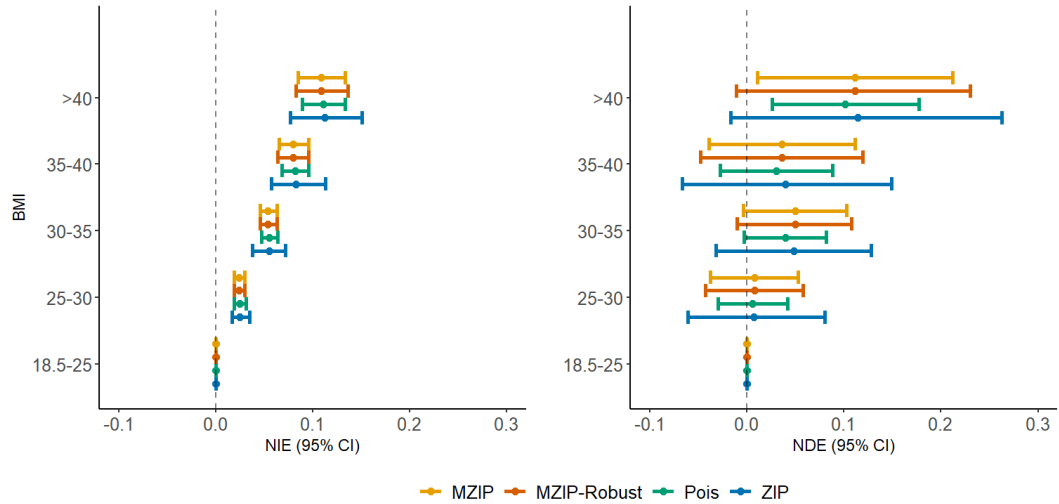

*Web Figure 2: Forest plot comparing NIE and NDE (Difference Scale) for Poisson vs. MZIP vs. ZIP in the data application using REGARDS.*

Supporting Information References